\documentclass[preprint,11pt,authoryear]{elsarticle}

\usepackage{amssymb}
\usepackage{amsmath}
\usepackage{setspace}
\usepackage{float}
\usepackage{algorithm}
\usepackage{algorithmic}
\usepackage{placeins}
\usepackage{graphicx}
\usepackage{subfigure}
\usepackage[table]{xcolor}
\usepackage{mathtools}
\usepackage{bbm}
\newtheorem{remark}{Remark}
\newtheorem{assumption}{Assumption}

\journal{European Journal of Operational Research}

\begin{document}

\begin{frontmatter}
\title{Mixed-integer optimization for multi-year military aircraft fleet management}

\author[1]{Matteo Vescovi \corref{cor1}}
\ead{matteo.vescovi@polimi.it}
\cortext[cor1]{Corresponding author}

\author[1]{Raffaele Giuseppe Cestari}
\author[3]{Ten. Col. Roberto Valdambrini}
\author[3]{Col. Andrea Mercurio}
\author[2]{Valentina Breschi,}
\author[1]{Mara Tanelli} 

\affiliation[1]{organization={Dipartimento di Elettronica, Informazione e Bioingegneria, Politecnico di Milano},
            addressline={Via Giuseppe Ponzio, 34}, 
            city={Milano},
            postcode={20133},
            country={Italy}}
\affiliation[2]{organization={Electrical Engineering Department, Eindhoven University of Technology},
            city={Eindhoven},
            postcode={5600MB},
            country={The Netherlands}}
\affiliation[3]{organization={Aeronautica Militare Italiana},
            city={Rome},
            country={Italy}}

\begin{abstract}
While existing strategies for Flight and Maintenance Planning for the defense sector generally address idealized conditions, real-world planning often involve non-nominal initial fleet states and complex inspection schemes. To address these challenges, we propose a multi-year planning strategy that maximizes long-term fleet availability. The formulation incorporates multiple competing objectives capturing the tight coupling between aircraft usage and maintenance, while enforcing cyclic inspection requirements and limited maintenance-dock capacity. To enhance practical applicability, we introduce three complementary strategies: a receding-horizon approach to reduce computational burden, a Bayesian Optimization routine to automatically tune cost-function coefficients, and an aircraft-specific weighting scheme that adapts model priorities based on fleet-condition statistics. Results on three realistic benchmark scenarios demonstrate the effectiveness of the approach. 
\end{abstract}

\end{frontmatter}

\section{Introduction}
\label{sec:introduction}
An effective Flight and Maintenance Planning (FMP) scheme should ensure long-term fleet efficiency by preventing inspection bottlenecks, maintaining high aircraft availability, minimizing maintenance resource waste, and maximizing aircraft utilization. Given the problem’s complexity and its importance for both civil and defense operations,  several approaches have been proposed over the last decades to tackle the FMP problem. 
Most of the early research on FMP focuses on tackling this problem for civil aircraft fleets, specifically looking at maintaining and planning for commercial passenger and cargo fleets. In these cases, aircraft are assigned to scheduled flight segments to build short-term itineraries, typically on a daily or weekly basis, while maintenance is generally limited to overnight checks~\citep{biro_civil,clarke_civil,gopalan_civil,moudani_civil,siriam_civil,cohn_civil}). In this context, the planning horizon is thus relatively short (often no more than a few days), and FMP formulations prioritize objectives such as profit maximization~\citep{yan_civil1,yan_civil2}), turnaround time reduction~\citep{ahire_civil}, and crew/workforce scheduling optimization~\citep{dijkstra_civil}. While these formulations are well-suited to tackle FMP for high-frequency, route-driven operations, they are instead not applicable for FMP in the defense sector, where $(i)$ the planning horizons are longer, $(ii)$ the maintenance policy depends on use, which may be irregular, $(iii)$ aircraft routing is not required, and $(iv)$ profit is not the main driver of the planning. 
Therefore, although civil-sector FMP remains the most extensively studied, fewer and more specialized works address the demands of mission-oriented aircraft planning. On one hand, institutional frameworks, such as the U.S. Army guidelines~\citep{usa_manual}, outline procedures for aircraft scheduling and inspection cycles in operational contexts. On the other hand, early research contributions include~\citet{sgaslik_militar}, who propose a helicopter-specific scheme built on two linked mixed-integer subproblems: one for inspections and training, and one for operational deployments. Similarly,~\citet{pippin_militar} presents linear and quadratic Mixed Integer Programming (MIP) problems to smooth Flight Hour (FH) distribution and maintain a steady flow of aircraft into phase maintenance; this formulation assumes a predefined maintenance plan and focuses solely on operational planning.
Instead,~\citet{radosavljevic_militar} targets narrower operational settings: they optimize fighter formation deployments against adversaries using fuzzy logic and integer programming.  \citet{kurokawa_militar} employ neural networks for air transportation planning for the Japan Air Self-Defense Force, though such methods may require large datasets that are rarely available. On a different note,~\citet{yeung_militar} propose a hybrid simulation–heuristic framework for mission assignment and maintenance scheduling in systems with multiple operational states, demonstrated via a hypothetical fleet; however, its reliance on simulated ideal conditions may limit robustness to unseen scenarios.
A different direction with respect to the aforementioned works is taken in~\citet{grecoMultiObj,FlottaGreca2}, where a mixed integer optimization problem is set up with the goal of maximizing aircraft availability while satisfying a set of constraints ensuring operational efficiency. This objective is attained by setting suitable bounds on the average number of available aircraft, the Residual Flight Hours (RFH), and the Residual Maintenance Time (RMT). With respect to maintenance, these first works consider a single maintenance station, whose capacity limits are imposed via structural constraints that ensure it is never empty when aircraft are awaiting inspection. Flight requirements are instead imposed via \textquotedblleft slacked\textquotedblright \ constraints
defined by squadron.
Building on these seminal works, \citet{FlottaGreca3} propose a related formulation that jointly maximizes each aircraft’s cumulative RFH and minimizes deviations from the fleet average, using the same constraints as~\citet{FlottaGreca2}. The key distinction is that~\citet{FlottaGreca3} enforce lower bounds on RFH and RMT per aircraft, whereas~\citet{FlottaGreca2} impose them as fleet-wide averages. The same authors revert to a single-objective formulation in \citet{FlottaGreca4}, aiming to maximize cumulative RFH across aircraft and periods. While retaining the constraints of~\citet{FlottaGreca2}, they add an upper bound on flight hours per aircraft per period; analogous bounds on RFH and RMT are omitted because they appear to lead to infeasibility problems.
\citet{Verhoeff} proposes an alternative single-objective formulation that maximizes the fleet’s \textquotedblleft minimum sustainability\textquotedblright (i.e., its RFH each period) while enforcing a lower bound on \textquotedblleft minimum serviceability\textquotedblright (i.e., the number of aircraft available for missions). A soft constraint also keeps availability in each period close to the required amount. In contrast,~\citet{NavalResearchGreco} introduce a bi-objective model that minimizes deviations between actual and target cumulative RFH and between target and actual RMT for each aircraft.
All these works treat FMP as \textquotedblleft deterministic\textquotedblright, ignoring unexpected events such as mission issues, component failures, or maintenance delays.~\citet{Mattila}, however, incorporate uncertainty through a disturbance component, minimizing expected unavailability and deviations in maintenance start times. Rather than modelling disturbances,~\citet{Safaei} addresses \textquotedblleft entry waves\textquotedblright (missions requiring multiple aircraft) by maximizing average fleet availability during each wave.
Because the previous approaches rely on mixed-integer programs, they become computationally expensive for large fleets. To mitigate this,~\citet{Peschiera} use Machine Learning to predict features of optimal MIP solutions and accelerate planning. Their method enriches standard branch-and-bound with ML-predicted cuts, substantially reducing solution times while maintaining accuracy and feasibility, and yielding more balanced maintenance schedules by lowering check-interval variance. Computational demands are even higher in long-term planning, which \citet{Peschiera2} address by proposing a heuristic that rapidly generates feasible solutions to warm-start the MIP solver.

\subsection{Contribution}
Building on previous works, we propose a new multi-objective, mixed integer problem for \emph{use-dependent} fleet and management planning. Each aircraft is associated with residual flight hours (FH) that decrease with usage, requiring inspections to be scheduled upon depletion. The objective is to reach a flight-hour targets for the entire fleet, while maintaining high availability by suitably scheduling the inspection scheme, under limited maintenance resources (e.g., available maintenance docks).
Unlike most existing approaches, the model captures multiple inspection types with different durations, triggered by usage, and organized in a predefined cyclic sequence. In particular, we consider four inspections performed at prescribed RFH intervals, with the last inspection of the cycle requiring specialized maintenance docks, further constraining scheduling decisions.
To address challenging initial fleet conditions and the computational burden of long-term multi-objective optimization, we introduce three practical enhancements. First, we adopt a \emph{receding horizon} strategy~\citep{rolling_horiz_ref1,rolling_horiz_ref2} to reduce computational cost while avoiding the myopic behaviour of solving a MIP separately for each year.
Second, by analysing some statistics of the fleet's initial condition, we propose an aircraft-specific weighting scheme that adjusts objective penalties; enabling \textquotedblleft informed\textquotedblright \  prioritization between flight-hour consumption and inspections.
Finally, we introduce a Bayesian Optimization routine (BO)~\citep{BO_ref} to automatically tune loss-function weights, improving solution quality while significantly reducing user intervention.
Results on three realistic benchmark fleets demonstrate the effectiveness and practical relevance of the proposed approach.

\subsection{Outline}
The paper is organized as follows. In Section~\ref{sec:targets_constraints}, we introduce the objectives that we aim to achieve with our planning scheme, as well as the constraints characterizing the FMP problem we aim to tackle. The mathematical translation of both these objectives and constraints is provided in Section~\ref{sec:mathematical_formulation}, where we introduce the overall formulation of our multi-year, multi-objective FMP problem. We then present the set of strategies proposed to solve the problem more effectively in practice in Section~\ref{sec:practical_implementation}. The results obtained on three benchmark case studies are presented and discussed in Section~\ref{sec:results}, and the paper then concludes with some final remarks and directions for future work.

\section{Introducing planning objectives and constraints}
\label{sec:targets_constraints}
We first introduce the objectives and constraints of the flight planning scheme, offering a \textquotedblleft natural language\textquotedblright interpretation before presenting their mathematical formalization. Nonetheless, this introduction already testifies to the interconnections between the flight and maintenance planning problem and the logistics of the involved maintenance infrastructures (i.e., the maintenance docks), as well as to the specific preferences due to internal procedures of the fleet's owner, spotlighting the intricacies of the FMP problem. 

\begin{remark}\label{remark:time}
    In our setting, objectives and constraints are enforced monthly.
\end{remark}

\subsection{Fleet planning goals}\label{sec:targets}

Designing a flight and maintenance plan for a military fleet requires balancing concurrent objectives. Although the fleet should reach a uniform distribution of residual flight hours across the fleet (i.e., reach a target \emph{flight hour scaling}, 
see~\figurename{~\ref{fig:scenarios}} in section \ref{sec:experiments};~\cite{FlottaGreca2,FlottaGreca4}) by the end of the planning horizon to guarantee readiness, pursuing this goal affects monthly flight hours allocation and maintenance schedule. Hence, rather than enforcing target matching directly, we guide the fleet toward it through a set of interconnected sub-objectives. Let us assume the following with respect to the available maintenance docks.
\begin{assumption}
    The number of maintenance docks is sized to match the needs of the fleet if all docks are occupied at all times
\end{assumption}
According to this assumption, in principle, no docks should, thus, be idle. Therefore, the first maintenance-related goal is the \emph{saturation of inspection docks}. If fulfilled, this objective makes docks (mostly) occupied at most (if not all) times, avoiding maintenance delays and guaranteeing smooth operations. 
In parallel, keeping aircraft idle is undesirable to meet operational needs. Thus, every aircraft not in maintenance should fly a user-defined minimum number of hours. Moreover, to support a balanced use of operational and maintenance resources, our second objective discourages a uneven FH consumption over the planning horizon, while a third objective accelerates it when an aircraft is close to an inspection. Although a few aircraft with low RFH are expected in use-based planning, having many in this condition could jeopardize readiness under unforeseen events.
Furthermore, typical fleet operation requires concentrating FH consumption in specific periods. As an example, the fourth objective can be weighted, allowing deviations favouring usage in the first part of the year.
Finally, fleets and squadrons are generally required to achieve predefined FH targets over the planning horizon. Meeting this fifth objective ensures operational duties are fulfilled, and due to its importance, it is enforced as a (possibly \textquotedblleft soft\textquotedblright) constraint in the FMP model.

\subsection{Fleet planning constraints}\label{sec:constraints}
In our setting, the fleet follows a standardized cyclic inspection scheme applied uniformly to all aircraft and repeated indefinitely. 
This scheme defines four maintenance types and the FH replenished upon completion the inspection; all share the same duration and restored FH except one, said \textquotedblleft main\textquotedblright \ or \textquotedblleft  major\textquotedblright \ inspection, which lasts longer, aggregates all other procedures, and can only be performed in specialized docks. Maintenance must respect dock capacity limits: each dock can service only one aircraft at a time, and each aircraft can occupy only one dock. Moreover, only a subset of docks is qualified for major inspections, further constraining their scheduling.
When not in inspection, stall periods are allowed only when the user-defined minimum flight hour threshold is zero; otherwise, aircraft must fly (up to a defined maximum threshold) to satisfy operational requirements. These bounds complement the fleet and squadron FH targets set by the planner. While these targets may be imposed as hard constraints, they can also be relaxed through soft constraints to retain feasibility in critical situations (i.e., allowing deviations from target FH when necessary). This requires \emph{slack} variables, which form our fifth optimization objective, minimized according to the desired strictness in enforcing FH targets.
Finally, the formulation includes appropriate initial and boundary conditions to ensure a complete and well-defined operational framework.

\section{Formulating the FMP problem}
\label{sec:mathematical_formulation}
We now formalize the \textquotedblleft natural language\textquotedblright \ goals and constraints introduced in Section~\ref{sec:targets_constraints} toward formulating the proposed FMP problem. To this end, let our fleet be composed of $N \in \mathbb{N}$ aircraft, partitioned into $F \in \mathbb{N}$ squadrons to which each aircraft is uniquely assigned, and let $C \in \mathbb{N}$ be the number of docks available for their maintenance. The considered planning horizon is $T$ monthly units.
By denoting with $h_{n,t,c} \in \{0,1\}$ the variable indicating if the $n$-th aircraft is undergoing a maintenance in the $c$-th dock ($h_{n,t,c}=1$) or not ($h_{n,t,c}=0$) at the $t$-th month, the first objective introduced in Section~\ref{sec:targets} can be written as follows:
    \begin{equation}\label{eq:j1}
    J_{1}(\mathcal{H})=\sum_{n=1}^{N}\sum_{t=1}^{T}\left(1-\sum_{c=1}^{C}h_{n,t,c}\right),
\end{equation}
with $\mathcal{H}=\{h_{n,t,c} \in \{0,1\} \mbox{~for all~} n \in [1,N], c \in [1,C], t \in [1,T]\}$. When minimized, this loss prevents the inspection dock from being idle as the least cost is attained when $\sum_{c=1}^{C}h_{n,t,c}=1$ and, thus, the $n$-th aircraft is undergoing an inspection. Note that, by further enforcing
\begin{equation}\label{eq:consistency1}
    0 \leq \sum_{c=1}^{C}h_{n,t,c} \leq 1,
\end{equation}
we further guarantee that each aircraft is assigned at most to one inspection dock when under maintenance.
Our second objective is to promote the monthly consumption of FH, which is achieved by additionally minimizing the loss
\begin{equation}
\label{eq:j2}
    J_{2}(\mathcal{X}) = \sum_{n=1}^{N}\sum_{t=1}^{T} (x_{n,t})^{2},
\end{equation}
with respect to the assigned flight hours of each aircraft at each month\footnote{$\mathbb{N}_{0}$ denotes the set of natural numbers including zero.}, i.e., $\mathcal{X}=\{x_{n,t} \in \mathbb{N}_{0} \mbox{~for all~} n \in [1,N], t \in [1,T]\}$. 
By using a quadratic loss in combination with a constraint on total flight hours (i.e., the flight target), this term further promotes a homogeneous distribution of flight hours.
Moreover, an additional cost is included to \textquotedblleft accelerate \textquotedblright \ FH consumption near maintenance, as aircraft with low residual flight hours pose a liability.
The latter is formulated as 
\begin{equation}\label{eq:j3}
    J_3(\mathcal{S},\mathcal{A},\mathcal{Z}) = \sum_{n=1}^{N}\sum_{t=1}^{T} a_{n,t}\left(r_n(s_{n,t})-z_{n,t}\right),
\end{equation}
where $r(s_{n,t}) \in \mathbb{N}_{0}$ indicates the amount of flight hours recovered by the $n$-th aircraft as a consequence of the last maintenance and $\mathcal{A}=\{a_{n,t} \in \{0,1\} \mbox{~for all~}n \in [1,N], t \in [1,T]\}$ is the set of aircraft availability indicators over the planning horizon, namely
\begin{equation}
    a_{n,t}=\begin{cases}
        0 \mbox{~if the~} n\mbox{-th aircraft is under inspection at the~} t\mbox{-th month},\\
        1 \mbox{~otherwise},
    \end{cases}
\end{equation}
for $t=[1,T]$ and $n=[1,N]$. Note that the contribution of non-available aircraft to the overall value of $J_3(\mathcal{R},\mathcal{A},\mathcal{Z})$ in~\eqref{eq:j3} is thus null. The former quantity $r(s_{n,t})$ is dictated by the variable
$s_{n,t} \in \{0,1,2,3\}$, which evolves as
\begin{equation}\label{eq:c2}
    s_{n,t+1} = s_{n,t} +(1-4v_{n,t})\sum_{c=1}^{C}d_{n,t,c},~~~n \in [1,N], t \in [1,T-1],
\end{equation}
thus increasing of one every time the $n$-th aircraft exists inspection and resetting to $0$ at the last major inspection thanks to
\begin{equation}\label{eq:auxiliary1}
    v_{n,t}=\begin{cases}
        1 \mbox{~if~} s_{n,t}=4,\\
        0 \mbox{~if~} s_{n,t} \in \{0,1,2,3\},
    \end{cases}
\end{equation}
for all $n \in [1,N]$ and $t \in [1,T]$. For this reason, we will refer to $s_{n,t}$ as the \textquotedblleft maintenance index\textquotedblright. Meanwhile, $\mathcal{Z}=\{z_{n,t} \in \mathbb{N}_{0} \mbox{~for all~}n \in [1,N], t \in [1,T]\}$ denotes the RFH to the next inspection across the fleet at all time instants, with $z_{n,t}$ evolving as
\begin{equation}\label{eq:c4}
    z_{n}(t+1)\!=\!z_{n,t}\!-\!x_{n,t}\!+\!\left(\sum_{c=1}^Cd_{n,t+1,c}\right)r_{n}(s_{n,t}),\,\,\forall n \in [1,N],~t \in [1,T-1].
\end{equation}
The variable $d_{n,t,c} \in \{0,1\}$ indicates wether the $n$-th aircraft is exiting an inspection undergone at the $c$-th dock at month $t$ (i.e., $d_{n,t,c}=1$) or not ($d_{n,t,c}=0$), with $n \in [1,N]$, $c \in [1,C]$ and $t \in [1,T]$, and it further satisfy
\begin{equation}\label{eq:consistency2}
    0 \leq \sum_{c=1}^{C} d_{n,t,c} \leq 1,~~\forall n \in [1,N],~t \in [1,T].
\end{equation}
so that each aircraft can exit maintenance from one and only one of the available docks. The dynamics in~\eqref{eq:c4} thus implies that the RFH are progressively reduced based on how much an aircraft is used (i.e., it flies), until its entrance in a maintenance dock, after which it is kept to zero thanks to the constraint
\begin{equation}\label{eq:logic1}
    z_{n,t} \leq a_{n,t} r(s_{n,t}),~~~n \in [1,N], t \in [1,T].
\end{equation}
Once the inspection is completed, the last term in~\eqref{eq:c4} guarantees that the residual flight hours to the next inspection are reset to the (non-null) value indicated by $r_{n}(t-1)$. 
Together with the losses in~\eqref{eq:j2}~-~\eqref{eq:j3}, a fourth cost function is introduced to shape flight consumption. Specifically, the following loss
\begin{equation}\label{eq:j4}
    J_{4}(\mathcal{X})=\sum_{n=1}^{N}\sum_{t=1}^{T}t^{\gamma}x_{n,t},
\end{equation}
penalizes differently the consumption of flight hours along the planning horizon, so that the closer to the end of it, the higher the cost (modulated via a tunable parameter $\gamma\geq 0$).
The last target that should be met is the achievement of the yearly target flight hours $x^{\mathrm{o}}_{f}$ that each squadron $f \in [1,F]$ should achieve at the end of the planning horizon. In our formulation, this constraint is enforced as follows
\begin{equation}\label{eq:fleetTarget}
    x^{\mathrm{o}}_{f}-\underline{\varepsilon}_{f}\leq \sum_{n=1}^{N}\sum_{t=1}^{T}x_{n,t}\mathbbm{1}(n,f)\leq x_{f}^{\mathrm{o}}+\overline{\varepsilon}_{f},~~f\in[1,F],
\end{equation}
thus allowing for eventual deviations from the target value, with $\mathbbm{1}(n,f)$ being an indicator function stating whether the $n$-th aircraft belongs to the $f$-th squadron, i.e.,
\begin{equation}
    \mathbbm{1}(n,f)=\begin{cases}
        1 \mbox{~if the~} n\mbox{-th aircraft belongs to the~} f\mbox{-th squadron},\\
        0 \mbox{~otherwise.}
    \end{cases}
\end{equation}
The entity of eventual departures from the target FH is \textquotedblleft controlled\textquotedblright \ via the two, in principle different, positive, slack variables $\underline{\varepsilon}_{f} \geq 0$ and $\overline{\varepsilon}_{f} \geq 0$, which can be different for each squadron $f \in [1,F]$, that are minimized via the loss
\begin{equation}\label{eq:j5}
    J_{5}(\mathcal{E})=\sum_{f=1}^{F}(\underline{\varepsilon}_{f}+\overline{\varepsilon}_{f}),
\end{equation}
where $\mathcal{E}=\{\underline{\varepsilon}_{f},\overline{\varepsilon}_f \in \mathbb{N}_0 \mbox{ for all } f \in [1,F]\}$. Apart from this constraint, the flight hours of each aircraft in each month should satisfy the constraint
\begin{equation}\label{eq:logic2}
        x_{n,t}\leq \min\left\{z_{n,t}, \,a_{n,t}\sum_{f=1}^{F}\Psi_f\mathbbm{1}(n,f)\right\},
\end{equation}
for all $n \in [1,N]$ and $t \in [1,T]$. This constraint enforces the flight hours not to exceed the RFH to the next inspection, while allowing aircraft to fly up to a maximum number of FH dictated by the constants $\{\Psi_f\}_{f=1}^{F}$ based on the squadron they belong to. The RFH to the next major inspection are denoted via the variable $y_{n,t}$ and evolve as
\begin{equation}\label{eq:c3}
    y_n(t+1)=  y_{n,t}-x_{n,t}+\bar{Y}v_{n,t}\sum_{c=1}^{c}d_{n,t,c},~~~n \in [1,N], t \in [1,T-1],
\end{equation}
with $v_{n,t}$ defined as in \eqref{eq:auxiliary1} and $\bar{Y}$ denoting the maximum RFH to the main inspection. 
In addition to these restrictions on flight hours, some constraints have to be imposed in relation to maintenance time and resources. To this end, let us denote the Residual Maintenance Time (RMT) of the $n$-th aircraft at the beginning of the $t$-th month with $g_{n,t} \in \mathbb{N}_0$. In our formulation, its evolution is dictated by
\begin{equation}\label{eq:c5}
    g_{n,t+1}=g_{n,t}+\sum_{c=1}^{C}\left[m_{n,t,c}G(s_{n,t})-h_{n,t,c}\right],~~~\forall n \in [1,N],~t \in [1,T-1], 
\end{equation}
so that the RMT value is restored to its maintenance-type dependent maximum value $G(s_{n,t-1})$ when a new inspection begins. This last aspect is encoded in the binary variable $m_{n,t,c} \in \{0,1\}$ for $c \in [1,C]$ which equals 1 when aircraft $n$ enters inspection at dock $c$ at time $t$, and 0 otherwise; while satisfying
\begin{equation}
    0 \leq \sum_{c=1}^{C}m_{n,t,c}\leq 1.
\end{equation}
Accordingly, the $n$-th aircraft can be associated with one and only one inspection dock. The variable $ g_{n,t}$ has also to satisfy the following inequalities
\begin{subequations}\label{eq:logic3}
    \begin{align}
        & g_{n,t}\leq (1-a_{n,t})G(s_{n,t}),~~n \in [1,N],~t \in [1,T],\\
        & \sum_{c=1}^{C} h_{n,t,c}\leq g_{n,t},~~n \in [1,N],~t \in [1,T],
    \end{align}
\end{subequations}
respectively ensuring each aircraft is not kept under maintenance for more than the number of months dictated by $G(s_{n,t})$, while being under inspection only when its RMT is higher than $0$. These conditions on the residual maintenance time have to be paired with constraints that allow the maintenance docks' capacity not to be exceeded. In particular, by indicating with $\Gamma$ the number of docks that are equipped to perform major inspections, we impose
\begin{subequations}\label{eq:c7}
    \begin{align}
        & \sum_{n=1}^{N}(1-a_{n,t})v_{n,t}\leq \Gamma,\\
        & \sum_{n=1}^{N}(1-a_{n,t})(1-v_{n,t})\leq C-\Gamma,
    \end{align}
at all $t \in [1,T]$. The first constraint guarantees that the number of aircraft undergoing a major inspection does not exceed the number of maintenance docks where such an inspection can be performed. Meanwhile, the second inequality enforces the number of aircraft undergoing a non-major inspection not to exceed the number of remaining inspection docks. 
\end{subequations}
Lastly, some auxiliary constraints are needed for the availability indicator $a_{n,t}$ to change consistently with the variables indicating the $n$-th aircraft entering or exiting maintenance. Such consistency is ensured by enforcing for $n \in [1,N],\,\, t \in [1,T]$ the following:

\begin{subequations}\label{eq:c6}
\begin{align}
    & \sum_{c=1}^{C} d_{n,t,c}\! =\! 1 
      \!\!\iff \!\! a_{n,t}=0 \land a_{n,t+1}=1, \\
    & \sum_{c=1}^{C} m_{n,t,c} \!=\! 1 
      \!\!\iff\!\! a_{n,t}\!=\!1 \land a_{n,t+1}\!=\!0 ,
\end{align}
while concurrently imposing 
\begin{equation}
    a_{n,t}=1 \iff g_{n,t}=0,~~~n \in [1,N], t \in [1,T], 
\end{equation}
so that an aircraft cannot be flagged as available if the associated RMT is non-null.
\end{subequations}

\subsection{FMP problem for defense aircraft fleets with cyclic maintenance}
\begin{table}[tb]
\caption{Set of user-defined constants characterizing the formulation of the FMP problem.}
    \label{tab:sets_constants}
    \centering
    \small
    \begin{tabular}{lll}
        \textbf{Constant} & \textbf{Values in} & \textbf{Meaning} \\
        \hline
         $r(s_{n,t})$ & $\mathbb{N}_0$ & FH recovered after the $s_{n,t}$-th inspection\\ 
         \hline
         $x_f^{\mathrm{o}}$ & $\mathbb{N}_0$ & Target FH for the $f$-th squadron\\
         \hline
         $\Psi_f$ & $\mathbb{N}_0$ & Max number of FH for the $f$-th squadron\\
         \hline
         $\bar{Y}$ & $\mathbb{N}_0$ & Max FH to the main inspection\\
         \hline
         $G(s_{n,t})$ & $\mathbb{N}_0$ & Duration of the $s_{n,t}$-th inspection\\
          \hline
          $\Gamma$ & $\{1,2,\ldots,C\}$ & Num. of docks capable of major inspections\\
          \hline
    \end{tabular}
\end{table}
\begin{table}[tb]
\caption{Optimization variables of the FMP problem and their meaning.\\ Indices ranges: $n \!\in\! [1,\!N], c\!\in\! [1,\!C], t \!\in\! [1,\!T], f \!\in\! [1,\!F]$. }
    \label{tab:sets_variables}
    \centering
    \small
    \begin{tabular}{lll}
        \textbf{Symbol} & \textbf{Meaning} & \textbf{Shorthand used in \eqref{eq:full_formulation}}\\
        \hline
        $h_{n,t,c}$ & Maintenance in the $c$-th dock & $\mathcal{H}\!=\!\{h_{n,t,c} \!\in\! \{0,\!1\}\}$\\
        \hline
        $x_{n,t}$ & FH & $\mathcal{X}\!=\!\{x_{n,t} \!\in\! \mathbb{N}_0 \}$\\
        \hline 
        $a_{n,t}$ & Aircraft availability & $\mathcal{A}\!=\!\{a_{n,t} \!\in\! \{0,\!1\} \}$\\
        \hline 
         $s_{n,t}$ & Next maintenance & $\mathcal{S}\!=\!\{s_{n,t} \!\in\! \{0,\!1,\!2,\!3\} \}$\\
         \hline 
         $v_{n,t}$ & Auxiliary variable & $\mathcal{V}\!=\!\{v_{n,t} \!\in\! \{0,\!1\} \}$\\
         \hline 
         $z_{n,t}$ & FH until next inspection & $\mathcal{Z}\!=\!\{z_{n,t} \!\in\! \mathbb{N}_0 \}$\\
         \hline 
         $d_{n,t,c}$ & Exit from the $c$-th dock & $\mathcal{D}\!=\!\{d_{n,t,c} \!\in\! \{0,\!1\} \}$\\
         \hline
         $\underline{\varepsilon}_f,\!\overline{\varepsilon}_f$ & Slack variables for squadron & $\mathcal{E}\!=\!\{\underline{\varepsilon}_f,\!\overline{\varepsilon}_f \!\in\! \mathbb{N}_0 \}$\\
         \hline 
         $y_{n,t}$ & FH until major inspection & $\mathcal{Y}\!=\!\{y_{n,t} \!\in\! \mathbb{N}_0 \}$\\
         \hline 
         $g_{n,t}$ & Residual Maintenance Time & $\mathcal{G}\!=\!\{g_{n,t} \!\in\! \mathbb{N}_0 \}$\\
         \hline 
         $m_{n,t,c}$ & Entrance in the $c$-th dock & $\mathcal{M}\!=\!\{m_{n,t,c} \!\in\! \{0,\!1\} \}$\\
         \hline
    \end{tabular}
\end{table}
Let us define $\mathcal{I}_N \coloneq [1,N]$, $\mathcal{I}_{T} \coloneq [1,T]$, $\mathcal{I}_{T-1} \coloneq [1,T-1]$; then, merging all the previous constraints and losses, assuming $n \in \mathcal{I}_n$, the overall FMP problem can thus be formulated as
\begin{subequations}\label{eq:full_formulation}
    \begin{align}        &\underset{{\substack{\mathcal{H},~\mathcal{X},~\mathcal{A},~\mathcal{Z},~\mathcal{E},~\mathcal{D},~\mathcal{S}\\
    \mathcal{M},~\mathcal{Y},~\mathcal{V},~\mathcal{G}}}}{\mathrm{minimize}} ~~J(\mathcal{H},\mathcal{X},\mathcal{S},\mathcal{A},\mathcal{Z},\mathcal{E})\\
    &\quad \mbox{s.t.}\,\,s_{n,t+1} \!=\! s_{n,t} \!+\!(1-4v_{n,t})\sum_{c=1}^{C}d_{n,t,c},\,\,t \in \mathcal{I}_{T},\label{eq:constraint1}\\
    &\quad  z_{n,t+1}(t+1)\!=\!z_{n,t}\!-\!x_{n,t}\!+\!\left(\sum_{c=1}^C d_{n,t+1,c}\right)r_{n}(s_{n,t}),\,\,t \in \mathcal{I}_{T},\\
    & \quad y_{n,t+1}\!=\!  y_{n,t}\!-\!x_{n,t}\!+\!\bar{Y}v_{n,t}\sum_{c=1}^{c}d_{n,t,c},\,\,t \in \mathcal{I}_{T},\\
    & \qquad  g_n(t+1)=g_{n,t}+\sum_{c=1}^{C}\left[m_{n,t,c}G(s_{n,t})-h_{n,t,c}\right],~~~ t \in \mathcal{I}_{T},\\
    & \qquad  v_{n,t}=1 \iff s_{n,t}=4,~~~t \in \mathcal{I}_{T},\\
    & \qquad    \sum_{c=1}^{C}d_{n,t,c}\!=\!1 \!\! \iff \!\! a_{n,t}\!=\!0 \land a_{n,t+1}=1,\,\,t \in \mathcal{I}_{T},\\
    & \qquad \sum_{c=1}^{C}m_{n,t,c}\!=\!1 \!\! \iff \!\! a_{n,t}\!=\!1\! \land\! a_{n,t+1}\!=\!0\! \land \!\sum_{i=1}^{t} x_n(j) \!=\! z_{n,t} ,\,t \in \mathcal{I}_{T},\\
    & \qquad  a_{n,t}=1 \iff g_{n,t}=0,~~~t \in \mathcal{I}_{T},\\
    &\qquad  x^{\mathrm{o}}_{f}-\underline{\varepsilon}_{f}\leq \sum_{n=1}^{N}\sum_{t=1}^{T}x_{n,t}\mathbbm{1}(n,f)\leq x_{f}^{\mathrm{o}}+\overline{\varepsilon}_{f},~~~\forall f\in[1,F],
    \end{align}
    \begin{align}
    &\qquad  x_{n,t}\leq \min\left\{z_{n,t},a_{n,t}\sum_{f=1}^{F}\Psi_f\mathbbm{1}(n,f),y_{n,t}\right\},\,\, t \in \mathcal{I}_{T},\\
    & \qquad  z_{n,t} \leq a_{n,t}r(s_{n,t}),~~~t \in \mathcal{I}_{T}, \\
    & \qquad  g_{n,t}\leq (1-a_{n,t})G(s_{n,t}),~~~t \in \mathcal{I}_{T},\\
    &\qquad  0 \leq \sum_{c=1}^{C}h_{n,t,c} \leq \min\{g_{n,t},1\},~~t \in \mathcal{I}_{T}\\
    & \qquad  0 \leq \sum_{c=1}^{C}d_{n,t,c} \leq 1,~ 0 \leq \sum_{c=1}^{C}m_{n,t,c} \leq 1,~~t \in \mathcal{I}_{T},\\
    & \qquad \sum_{n=1}^{N}(1\!-\!a_{n,t})v_{n,t}\leq \Gamma,\,\sum_{n=1}^{N}(1\!-\!a_{n,t})(1\!-\!v_{n,t})\leq C\!-\!\Gamma,~~t \in \mathcal{I}_{T},\label{eq:constraint_end}
    \end{align}
    where the overall cost $J = J(\mathcal{X},\mathcal{R},\mathcal{A},\mathcal{Z},\mathcal{E})$ is the weighted combination of the losses introduced in \eqref{eq:j1}-\eqref{eq:j3}, \eqref{eq:j4} and \eqref{eq:j5}, i.e.,
    \begin{equation}\label{eq:cost_funct}
    J \!= \!w_1J_1(\mathcal{H}) \!+\!w_2J_2(\mathcal{X}) \!+\!w_3J_3(\mathcal{S},\mathcal{A},\mathcal{Z}) \!+\!w_4J_4(\mathcal{X}) \!+\! w_5J_5(\mathcal{E}).
    \end{equation}
\end{subequations}

\section{Solution enhancing strategies}
\label{sec:practical_implementation}
We now discuss a set of strategies intended to practically handle the computational complexity of the fleet planning problem introduced in \eqref{eq:full_formulation}, as well as reducing the tuning burden at the user end and coping with \textquotedblleft unfavourable\textquotedblright \ initial fleet conditions.   

\subsection{Coping with the computational burden: receding horizon}\label{sec:rh}
\begin{figure}[]
\centering
\includegraphics[width=1 \columnwidth]{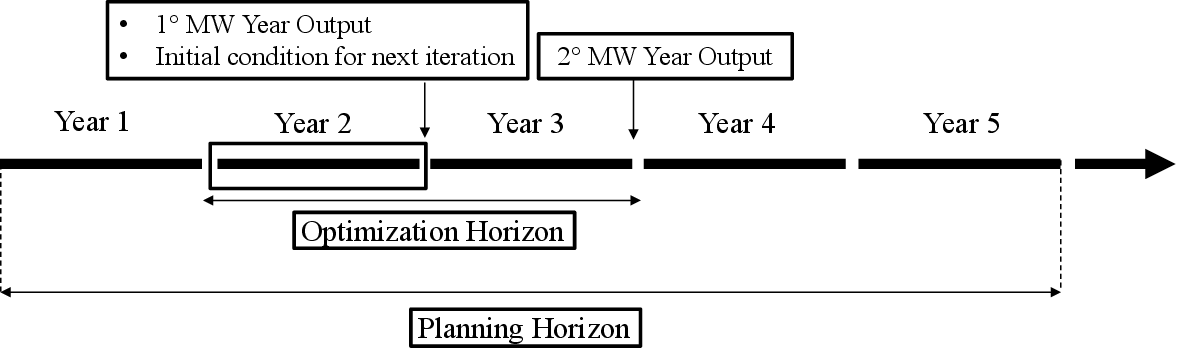}
\caption{A schematic representation of the receding horizon strategy over a planning horizon $T$ of five years, considering an optimization horizon of $2$ years. In this representation, MW stands for moving window.}\label{fig:MW}
\end{figure}

The complexity of the optimization problem in \eqref{eq:full_formulation} increases with the length of the planning horizon, due to the growing number of variables and constraints. In turn, this is likely to make the computational complexity and time needed to solve \eqref{eq:full_formulation} incompatible with practical needs, especially when the considered fleet is large.   
While solving the multi-year planning problem in one shot would be best, to cope with the aforementioned practical issues, we propose to exploit a \emph{receding horizon strategy}. As summarized in \figurename{~\ref{fig:MW}}, we thus consider a multi-year optimization horizon $L$ smaller than the actual planning one, i.e., $L<T$. 
At the start of each year, we solve a simplified version of~\eqref{eq:full_formulation} over a window of length $L$. Only the decisions for the first year are retained, while the remaining years are used to warm-start subsequent optimizations. The optimization window then slides one year ahead, and the procedure is repeated.
Note that, considering $L \neq 1$ and using the part of the solution of one \textquotedblleft reduced\textquotedblright \ problem to initialize the next allows us to \emph{leave the fleet in a favourable condition}, by favouring smooth transitions of FH consumption and aircraft's planned inspections over the years. 
At the same time, we acknowledge that this approach is suboptimal compared to solving the FMP problem over the full horizon. This optimality gap can be reduced by increasing $L$, at the cost of higher computational complexity.

\subsection{Bayesian optimization for weight tuning}\label{sec:BO_hyperparameter}
When looking at the FMP formulation in \eqref{eq:full_formulation}, it is clear that the main burden on the final user lies in the choice of the weights $\mathcal{W}=\{w_i\}_{i=1}^{5}$. Indeed, achieving a trade-off between the five objectives in \eqref{eq:cost_funct} is all but trivial even for expert users, given their heavily intertwined nature. At the same time, the main targets of a practitioner when devising a flight and maintenance planning scheme are twofold. 
On the one hand, the scaling of each squadron in the fleet should be as close as possible to the \emph{ideal} scaling (see \figurename{~\ref{fig:scenarios}}) constructed by a thumb rule generally accepted in the literature (see, e.g.,~\cite{FlottaGreca2}), i.e.,

\begin{equation}
    y^{\mathrm{o}}_{n,\phi+1}=\max\left\{0\,,\bar{Y}\frac{p_n(y_{n,12\phi})-N_{f}^{down}}{N_f -N_{f}^{down}}   \right\}
\end{equation}
where $N_f$ if the number of aircraft belonging to the $f$-th squadron and $N_{f}^{down}$ the average number of aircrafts under inspection\footnote{We compute it through a rule of thumb: $N_{f}^{down}=\lceil\frac{x_f^{\mathrm{o}}}{\bar{Y}}\rceil$}. $p_n(y_{n,12\phi})$ indicates the position of the $n$-th aircraft in a ranking constructed based on its distance from the major inspection relative to other aircrafts in its squadron, at the beginning of each year in the planning horizon. On the other hand, inspection docks should not be empty to avoid resource waste. 
The former goals can be translated into the following losses:
\begin{equation}\label{eq:scaling_loss}
    J_{\mathrm{flow}}(\mathcal{W})\!=\! \sum_{\phi=0}^{4}\sum_{n=1}^{N}\!|y_{n,\phi+1}^{\mathrm{o}}\!-\!y_{n,12(\phi+1)}(\mathcal{W})|,
\end{equation}
and
\begin{equation}\label{eq:dock_loss}
    J_{\mathrm{docks}}(\mathcal{W})=\sum_{t=1}^{T}\sum_{n=1}^{N}\sum_{c=1}^{C}(1-h_{n,t,c}(\mathcal{W})),
\end{equation}
in turn depending on the tunable weights in \eqref{eq:cost_funct} thanks to their dependence on $y_{n,12(\phi+1)}(\mathcal{W})$ and $h_{n,t,c}(\mathcal{W})$, which are obtained by solving the FMP problem. These two losses can then be merged into a unique cost encompassing both the objectives sought by a practitioner, i.e.,
\begin{equation}\label{eq:tuning_loss}
        J_{\mathrm{tune}}(\mathcal{W})=J_{\mathrm{flow}}(\mathcal{W})+\lambda J_{\mathrm{docks}}(\mathcal{W}),
\end{equation}
where $\lambda$ is a scaling factor used to make the scaling and dock-related losses comparable\footnote{in our tests we always set $\lambda=10^{4}$.}. 
This merged cost can then be used to guide the selection of the weights by solving the following (nested) optimization problem
\begin{equation}\label{eq:nested_problem}
    \begin{aligned}
        & \underset{\mathcal{W} \subseteq \Omega \in \mathbb{R}^{5}}{\mbox{minimize}} ~~J_{\mathrm{tune}}(\mathcal{W})\\
        &\qquad ~~\mbox{s.t.}~~\mathcal{Y}^{\star}(\mathcal{W}),\mathcal{H}^{\star}(\mathcal{W}) \in \arg\min ~~J(\mathcal{H},\mathcal{X},\mathcal{S},\mathcal{A},\mathcal{Z},\mathcal{E};\mathcal{W}) \\
        & \qquad \qquad \qquad \qquad \qquad \qquad \qquad \mbox{s.t.~}~ \eqref{eq:constraint1}-\eqref{eq:constraint_end},
    \end{aligned}
\end{equation}
where $\Omega$ is a compact set defined by the user to limit the search space for the cost penalties. Since finding a closed-form solution for this problem is not viable, we propose to use auto-tuning techniques often employed in the machine learning literature and beyond (see~\cite{bemporad_bo_ref,BO_ref}) to automate the tuning of $\{w_i\}_{i=1}^{5}$, specifically focusing ob Bayesian Optimization (BO) (see, e.g.,~\cite{bernardo2009bayesian,pelikan2005bayesian,victoria2021automatic}).
\begin{algorithm}[!tb]
\caption{Penalty Tuning via Bayesian Optimization}\label{alg:bo}
\small
\begin{algorithmic}
\STATE \textbf{Input:} Search space $\Omega$
\STATE \textbf{Initialize:} $D_{\mathcal{W}}=\{\mathcal{W}_{l}^{(0)},J_{\mathrm{tune}}(\mathcal{W}_l^{(0)})\}_{j=1}^{\Lambda}$
\FOR{$i = 0,...$}
\STATE $\mathcal{W}^{(i+1)} \leftarrow \arg\max_{\mathcal{W}}~\alpha(\rho(J_{\mathrm{{tune}}}(\mathcal{W})|D_{\mathcal{W}}))$
    \STATE Solve \eqref{eq:full_formulation} minimizing $J(\mathcal{H},\mathcal{X},\mathcal{S},\mathcal{A},\mathcal{Z},\mathcal{E};\mathcal{W}^{(i+1)})$
    \STATE Compute $J_{\mathrm{tune}}(\mathcal{W}^{(i+1)})$
    \STATE $D_{\mathcal{W}} \leftarrow D_{\mathcal{W}} ~\cup~(\mathcal{W}^{(i+1)},J_{\mathrm{tune}}(\mathcal{W}^{(i+1)})) $
    \STATE Update the surrogate model $\rho(J_{\mathrm{tune}}(\mathcal{W})|D_{\mathcal{W}})$
    \STATE Until a convergence criterion is satisfied
\ENDFOR
\STATE \textbf{Output:} Optimal penalties $\mathcal{W}^{\star}$
\end{algorithmic}
\end{algorithm}
BO is used to tackle the outer level of the nested problem \eqref{eq:nested_problem} through the iterative procedure summarized in Algorithm~\ref{alg:bo}. 
At a glance, starting from a set of randomly sampled weights and corresponding loss evaluations, a surrogate probabilistic model of $J_{\mathrm{{tune}}}(\mathcal{W})$ is built and updated at each iteration. New candidate weights are selected by optimizing a statistic derived from the surrogate (i.e., the acquisition function $\alpha$), evaluated by solving the FMP problem, and then added to the dataset. This process is repeated until a stopping criterion is met (e.g., a given number of iterations).
Beyond reducing manual hyperparameter tuning and making the FMP approach accessible to non-expert users, Algorithm~\ref{alg:bo} helps tackling inconvenient initial fleet conditions.

\subsection{Hierarchical scheme for unbalanced initial fleet conditions}
\label{sec:hierarchy}
When the initial conditions of the fleet are particularly critical, we propose to further manipulate the weights of the loss in \eqref{eq:cost_funct}, by assigning the penalties for the different aircraft based on some statistics related to the initial condition. More specifically, we propose to modify the weights $w_1$, $w_2$, and $w_3$, making them aircraft-specific, by looking at the distances from the target scaling 
(see~\figurename{~\ref{fig:scenarios}}) as well as those from the next maintenance as follows:
\begin{subequations}\label{eq:weights}
    \begin{align}
        & w_{1,n}(\phi+1)=\bar{w}_1+\beta_1\sum_{f=1}^{F}\mathbbm{1}(n,f)(N_f-p_n(y_n(12\phi))),\label{eq:w0}\\
        & w_{2,n}(\phi+1)=\bar{w}_2-\beta_2\Delta_n(12\phi),\label{eq:w1}\\
        &w_{3,n}(\phi+1)=\bar{w}_3+\beta_3z_n(12\phi),\label{eq:w2}
    \end{align}
\end{subequations}
where $\Delta_n(12\phi)$ is the distance of the $n$-th aircraft with respect to the target scaling of its squadron at the beginning of each year $\phi+1$ in the planning horizon, with $\phi \in \{0,1,2,3,4\}$. 
These choices collectively shape aircraft prioritization and flight-hour consumption. Selecting $w_{1,n}(\phi+1)$ as in~\eqref{eq:w0} establishes a hierarchy based on scaling positions, penalizing aircraft closer to major maintenance and thus favoring their earlier entry into inspection. The choice of $w_{2,n}(\phi+1)$ in~\eqref{eq:w1} adapts the penalty on flight-hour consumption according to each aircraft’s deviation from the target scaling: FH use is encouraged when $\Delta_n>0$, preventing underutilization, and discouraged when $\Delta_n<0$, limiting overuse and mitigating maintenance queues. Finally, the penalty in~\eqref{eq:w2} further discourages FH consumption for aircraft nearing their next inspection, effectively accelerating their transition into maintenance.
These definitions introduce a squadron-level hierarchy that prioritizes some aircraft for flight or for maintenance, while requiring the selection of positive parameters $\{\bar{w}_i,\beta_i\}_{i=1}^{3}$. These can be chosen manually, via rescaling strategies, or automatically using the BO-based tuning method described in Section~\ref{sec:BO_hyperparameter}.
By correcting the weights in this aircraft-specific way, the loss $J=J(\mathcal{H},\mathcal{X},\mathcal{S},\mathcal{A},\mathcal{Z},\mathcal{E})$ in \eqref{eq:cost_funct} thus becomes:
\begin{align}\label{eq:hierarc_cost}
\nonumber J=\!\! \sum_{n=1}^{N}\sum_{t=1}^{T}\sum_{\phi=0}^{4}&\Biggl[w_{1,n}(\phi+1)\!\left(\!1\!-\!\!\sum_{c=1}^{C}h_{n,t,c}\right)\!+\!w_{2,n}(\phi+1)(x_{n,t})^2\!+\\
&\!w_{3,n}(\phi+1)(r(s_{n,t})a_{n,t}\!-\!z_{n,t})\Biggr]\!+w_4J_4(\mathcal{X})+ w_5J_5(\mathcal{E}),
\end{align}
where the last two elements are not modified with respect to the loss \textquotedblleft without hierarchy\textquotedblright \ already introduced in \eqref{eq:cost_funct}.

\section{Experimental results}\label{sec:experiments}
\label{sec:results}
We now showcase the performance of the proposed FMP scheme on three benchmark examples, showcasing the benefits of the strategies proposed to enhance its practicality.
Before describing the considered examples, the indicators used to quantitatively evaluate performance and the results we achieve, we recall that we consider a planning horizon of $5$ years (see also Remark~\ref{remark:time}). Since the complexity of the planning problem exceeds what can be handled with the employed device, a one-shot solution for all 5 years is, however, not practically computable with our hardware. Therefore, we present the results by considering two simplified optimization routines. On the one hand, we show what happens if the FMP problem is solved \textquotedblleft year-by-year\textquotedblright, i.e., we solve the problem for each year separately using the outcome of the FMP at a certain year as the initialization for the next. On the other hand, we showcase what happens by employing the receding horizon strategy introduced in Section~\ref{sec:rh} by fixing $L=24$, i.e., considering a prediction horizon of two years.
\begin{figure}[!tb]
\caption{Benchmark fleets: initial scaling \emph{vs} ideal one. The ideal scaling is indicated by the red line.}
	\label{fig:scenarios}
	\centering
        \begin{tabular}{ccc}
             \subfigure[Standard fleet \label{fig:scenario_good}]{\includegraphics[width=0.3\textwidth]{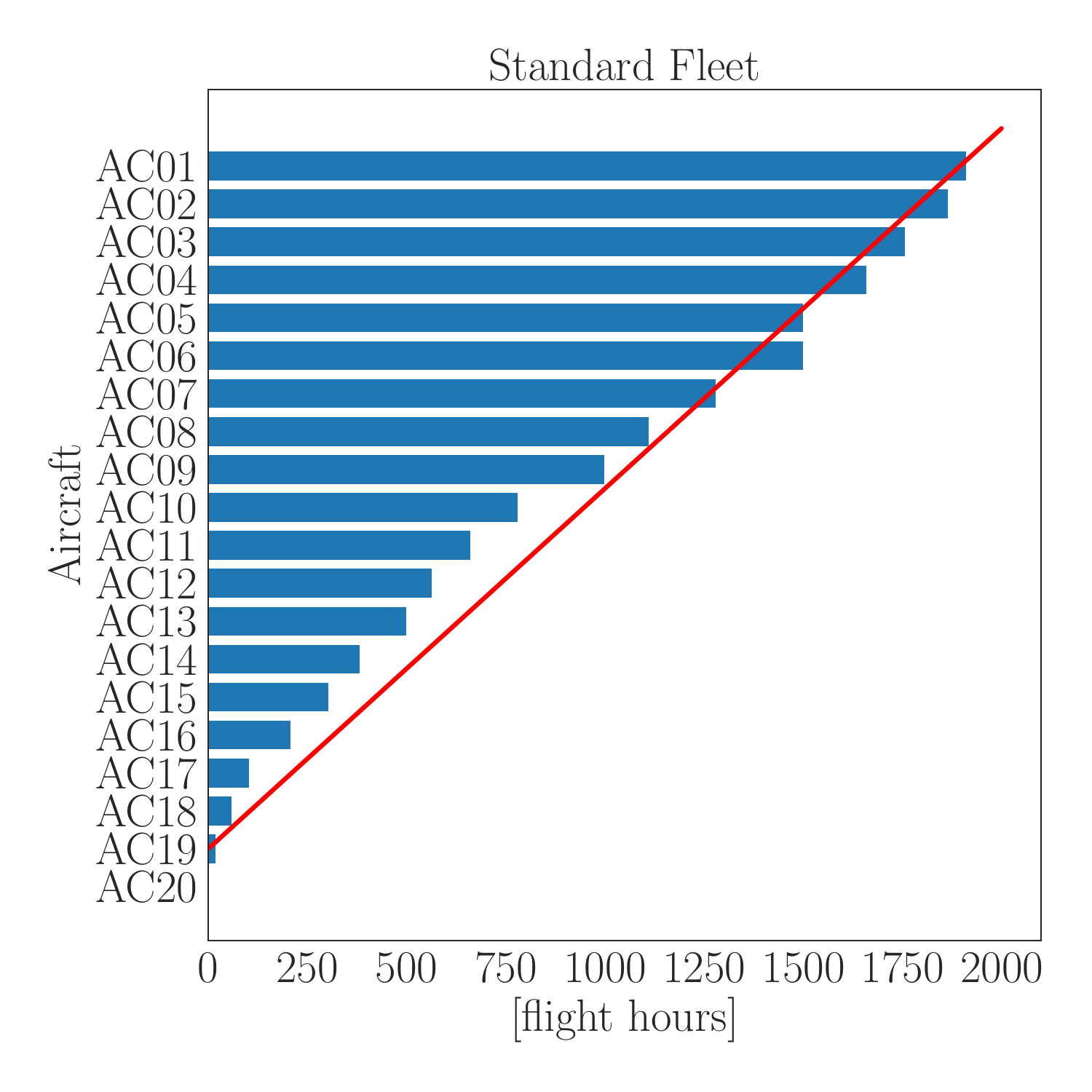}} & \subfigure[Critical fleet \label{fig:scenario_critical}]{\includegraphics[width=0.3\textwidth]{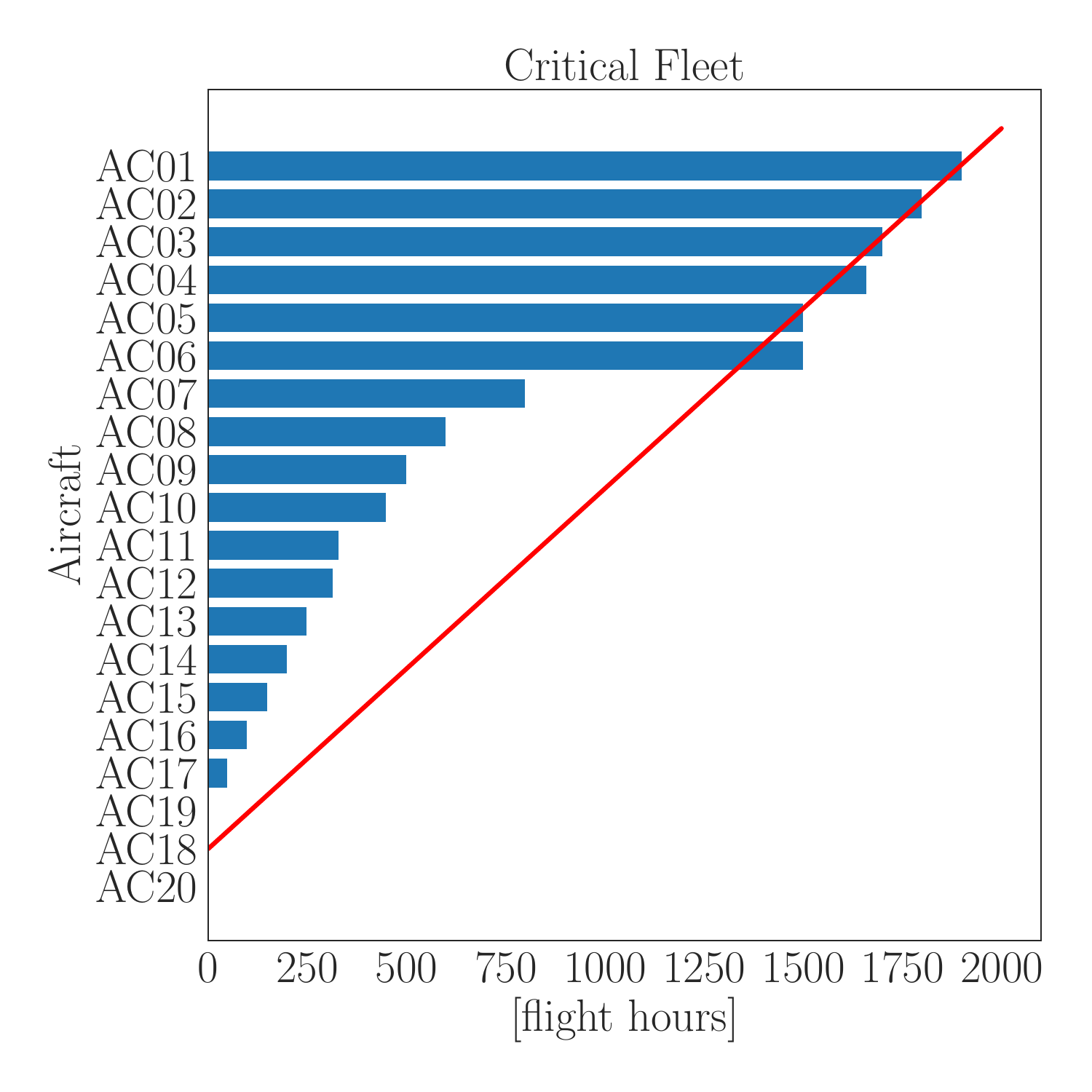}} & \subfigure[New fleet \label{fig:scenario_new}]{\includegraphics[width=0.3\textwidth]{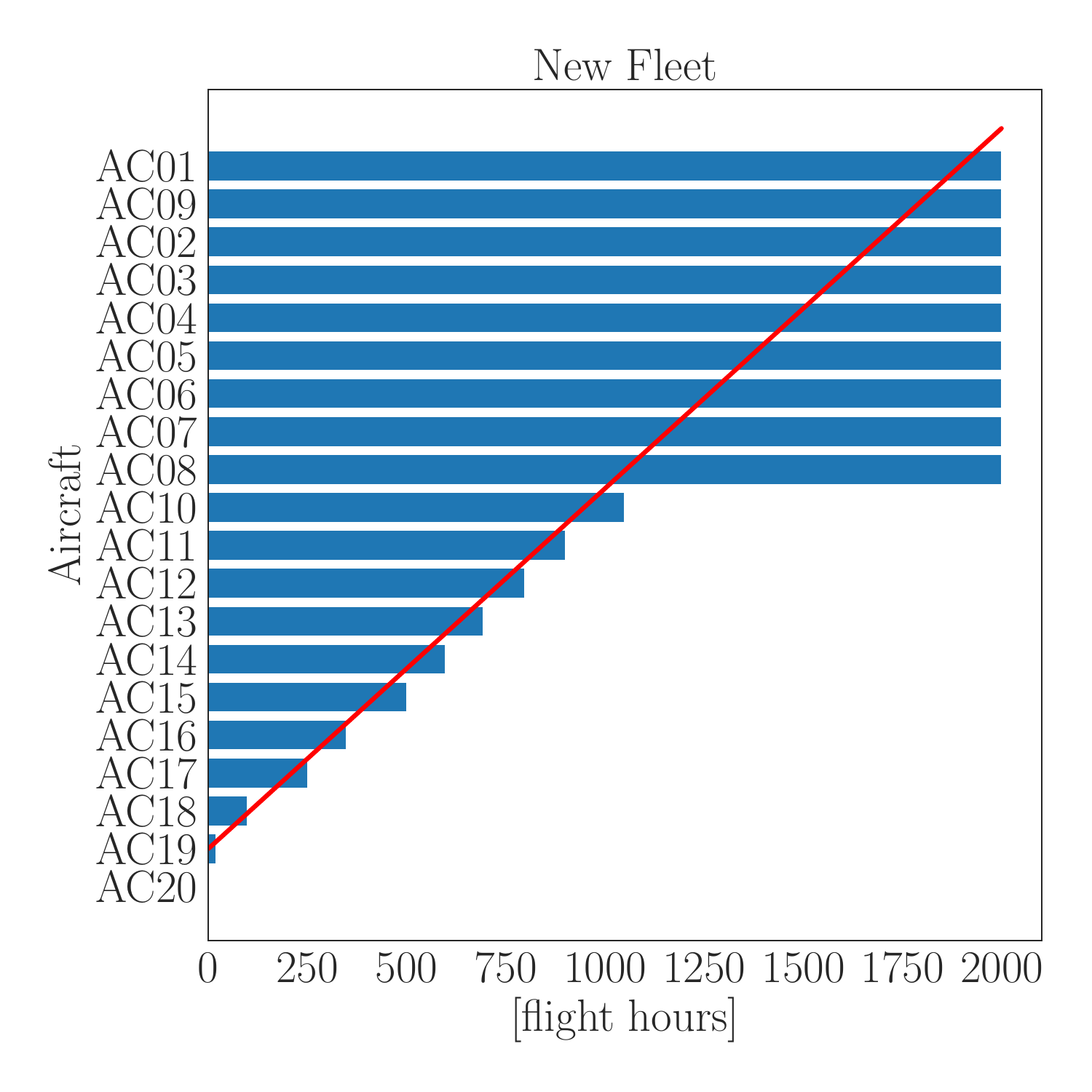}}  
        \end{tabular}
	
\end{figure}
\begin{table}[]
\centering
\small
\caption{Constants characterizing our FMP problem.}
\label{tab:paramteres}
\begin{tabular}{|l|l|l|l|l|l|l|}
\hline
symbol & $r(s_{n,t})$ & $x_1^{\mathrm{o}}$ & $x_2^{\mathrm{o}}$ & $\Psi_1$ & $\Psi_2$ & $\bar{Y}$ \\ \hline
value  & 500          & 3500               & 700                & 25       & 25       & 2000      \\ \hline
\end{tabular}
\end{table}
To evaluate the planning resulting from the proposed FMP scheme, we consider three benchmark fleet, all composed of $N=25$ aircraft organized in $F=2$ squadrons, respectively made of $20$ and $5$ aircraft. 
The fleet gets maintained in $C=5$ docks, of which $\Gamma=2$ are equipped to perform major inspections.
The initial scaling of the larger squadron in these fleets is shown in \figurename{~\ref{fig:scenarios}}, highlighting the core differences between the three considered planning scenarios. The fleet whose initial condition is reported in \figurename{~\ref{fig:scenario_good}} already has a scaling that is close to the ideal one. For this fleet, which we refer to as the \emph{standard fleet}, our goal is thus to reduce or maintain the distance to the ideal scaling characterizing the initial condition over the planning horizon. The second benchmark, denoted as the \emph{critical fleet}, is characterized by a scarcity of RFH left across the fleet. This feature ultimately leads to an initial scaling that is distant from the ideal one (see \figurename{~\ref{fig:scenario_critical}}), a situation from which the fleet has to recover through the proposed FMP scheme and the practical strategies introduced in Section~\ref{sec:practical_implementation}. The last fleet (see \figurename{~\ref{fig:scenario_new}}) includes a subset of new aircraft. All the latter have the maximum RFH possible, creating a \textquotedblleft wall\textquotedblright \ in the scaling, which might cause inspection queues if not smoothed out throughout the planning horizon.     
Additional features of the considered FMP problem are summarized in \tablename{~\ref{tab:paramteres}}, while the duration of minor (i.e., $s_{n,t}\neq 3$) and major ($s_{n,t}=3$) inspections are respectively set to 6 and 8. 

\subsection{Performance indicators}\label{sec:performance_indicators}
To evaluate the performance achieved with the proposed FMP scheme, we introduce a set of key performance indicators (KPIs) inspired by literature examples and discussion with Air Force operators.
First of all, we consider the number of \emph{Residual Flight Hours} to the next \emph{major inspection} at the end of each year, i.e.,
\begin{equation}\label{eq:a_RFH}
    \bar{y}(\phi+1)=\frac{1}{N}\sum_{n=1}^{N}(\bar{Y}-y_n(12\phi+12)),~~\phi=\{0,1,2,3,4\}.
\end{equation}
Indeed, the lower $\bar{y}(\phi+1)$ is, the more aircraft will be close to a major maintenance the next year, in turn causing inspection queues and possible delays in accomplishing mission requirements. We then exploit the yearly \emph{scaling loss} 
\begin{equation}\label{eq:scaling_year}
 J_{\mathrm{flow}}(\phi\!+\!1)\!=\!  \frac{\sum_{f=1}^{F}\sum_{n=1}^{N}\!\mathbbm{1}(n,f)}{N} \sum_{f=1}^{F}\sum_{n=1}^{N}\!\mathbbm{1}(n,f)|y_{n,12\phi\!+\!1}^{\mathrm{o}}-y_{n,12\phi+12}|, 
\end{equation}
already exploited in \eqref{eq:scaling_loss} to evaluate the goodness of the scaling of the RFH to the main inspection when compared to the target RFH. The distance of the fleet to the ideal scaling is further assessed via the \emph{FH Surplus} ($FHS$) and the \emph{FH Deficit} ($FHD$) for each fleet, respectively defined as
\begin{subequations}
    \begin{align}
        & \mathrm{FHS}_{f}(\phi\!+\!1)\!=\!\begin{cases}
            \frac{\sum_{n=1}^{N}\mathbbm{1}(n,f)\left[y_n^{\mathrm{o}}(\phi+1)-y_n(12\phi+12)\right]}{\sum_{n=1}^{N}\mathbbm{1}(n,f)y_n^{\mathrm{o}}(\phi+1)}, \mbox{ if } K_1>0,\\
            0 \mbox{ otherwise,}
        \end{cases}\\
        & \mathrm{FHD}_f(\phi\!+\!1)\!=\!\begin{cases}
             \frac{\sum_{n=1}^{N}\mathbbm{1}(n,f)\left|y_n^{\mathrm{o}}(\phi+1)-y_n(12\phi+12)\right|}{\sum_{n=1}^{N}\mathbbm{1}(n,f)y_n^{\mathrm{o}}(\phi+1)}, \mbox{ if } K_1<0,\\
            0 \mbox{ otherwise,}
        \end{cases} 
    \end{align}
\end{subequations}
with $K_1 = y_n(12\phi+12)-y_n^{\mathrm{o}}(12\phi+1)$, $f \in [1,F]$ and $\phi \in \{0,1,2,3,4\}$, reflecting excesses of FH to be consumed or deficits of FH. Both a surplus and a deficit of FH are undesirable from a planning perspective, as they are indicators of potential inspection queues in the successive planning years.    
Apart from these indicators, which are more focused on the operational readiness of the fleet, we also introduce a KPI explicitly linked to maintenance. Specifically, as successful planning leads to inspection docks that are not idle, we consider the number of \emph{Empty Docks} (ED) as an indicator of the quality of the planning from the maintenance perspective. It is worth remarking that having docks idle might undermine the operational readiness of the fleet, thus compromising also its operational effectiveness. 

\subsection{Receding horizon \emph{vs} year-by-year planning}\label{sec:receding_vs_yearly}
\begin{table}[!tb]
    \caption{Receding horizon \emph{vs} year-by-year: parameters in \eqref{eq:cost_funct}.}
    \label{tab:weigts_fixed}
    \centering
    \small
    \begin{tabular}{cccccc}
         $w_1$& $w_2$ & $w_3$ & $w_4$ & $w_5$ & $\gamma$ in \eqref{eq:j4}\\
         \hline
         0.8 & 0.15 & 0.2 & 0.15 & $10^{5}$ & 1.3\\
         \hline
    \end{tabular}
\end{table}
\begin{table}[!tb]
    \caption{Year-by-year \emph{vs} RH: KPIs for the standard fleet, where ED$(\phi+1)$ is the counter of yearly empty docks.}
    \label{tab:KPI_goodFleet}
    \centering
    \begin{tabular}{lccccccccc}
         \multicolumn{1}{c}{} & \multicolumn{2}{c}{$\bar{y}(\phi+1)$} & \multicolumn{2}{c}{$J_{\mathrm{flow}}(\phi+1)$} & \multicolumn{2}{c}{ED$(\phi+1)$} \\
         \hline
         \multicolumn{1}{c}{} year & year-by-year & RH & year-by-year & RH & {year-by-year} & RH \\
         \hline 
         $\phi=0$ & 190.56 & 179.76 & 308.58 & 283.82 & \cellcolor{green!5!white}0 & \cellcolor{green!5!white}0\\
         \hline 
         $\phi=1$ & 214.56 & 210.52 & 190.52 & 166.96 & \cellcolor{green!5!white}0 & \cellcolor{green!5!white}0\\
         \hline 
         $\phi=2$ & 217.44 & 241.32 & 268.80 & 219.01 & \cellcolor{yellow!5!white}5 &\cellcolor{green!5!white}0\\
         \hline 
         $\phi=3$& 238.84 & 232.60 & 294.11 & 119.78 & \cellcolor{yellow!5!white}5& \cellcolor{green!5!white}0\\
         \hline 
         $\phi=4$& \cellcolor{red!5!white}- & 266.64 & \cellcolor{red!5!white}- & 220.80 & \cellcolor{red!5!white}-&\cellcolor{green!5!white}2\\
         \hline 
    \end{tabular}
\end{table}
By fixing the weights characterizing \eqref{eq:cost_funct} as in \tablename{~\ref{tab:weigts_fixed}}, thus imposing a higher relative weight to idle maintenance docks, we first compare the results achieved with a (eventually myopic) year-by-year planning with the receding horizon strategy presented in Section~\ref{sec:rh}. As already mentioned, the prediction horizon is chosen as $L=24$, so that a window of two years is jointly planned. Some KPIs achieved for the standard benchmark fleet previously introduced are reported in \tablename{~\ref{tab:KPI_goodFleet}}.
Note that total FH are not reported, as the target is \emph{always} met across all configurations and planning years, except when the FMP problem is infeasible.
Results show that the RH strategy is beneficial. While a myopic year-by-year planning becomes infeasible despite favourable initial scaling, RH enables feasible planning over the full five-year horizon. RH also improves performance by increasing average yearly RFH and reducing empty docks, though gains are less pronounced given the fleet’s favourable initial conditions. 

 \subsection{Focusing on the critical fleet: the impact of Bayesian Optimization}
\label{sec:results_BO}
\begin{table}[!tb]
    \caption{Manual tuning \emph{vs} BO: parameters in \eqref{eq:weights}.}
    \label{tab:weights_critical_bo}
    \centering
    \small
    \begin{tabular}{l|l|l|l|l|l|l|}
\cline{2-7}
                             & $w_1$ & $w_2$ & $w_3$ & $\beta_1$ & $\beta_2$ & $\beta_3$ \\ \hline
\multicolumn{1}{|l|}{Manual} & 0.4   & 0.5   & 0.01  & 0.03      & 0.001     & 0.001     \\ \hline
\multicolumn{1}{|l|}{BO}     & 0.924 & 0.074 & 0.001 & 0.042     & 0.0007    & 0.044     \\ \hline
\end{tabular}
\end{table}

\begin{figure}[!tb]
    \centering
    \begin{tabular}{cc}
        \subfigure[Empty docks: manual \emph{vs} BO-tuning]{\includegraphics[width=0.4\textwidth]{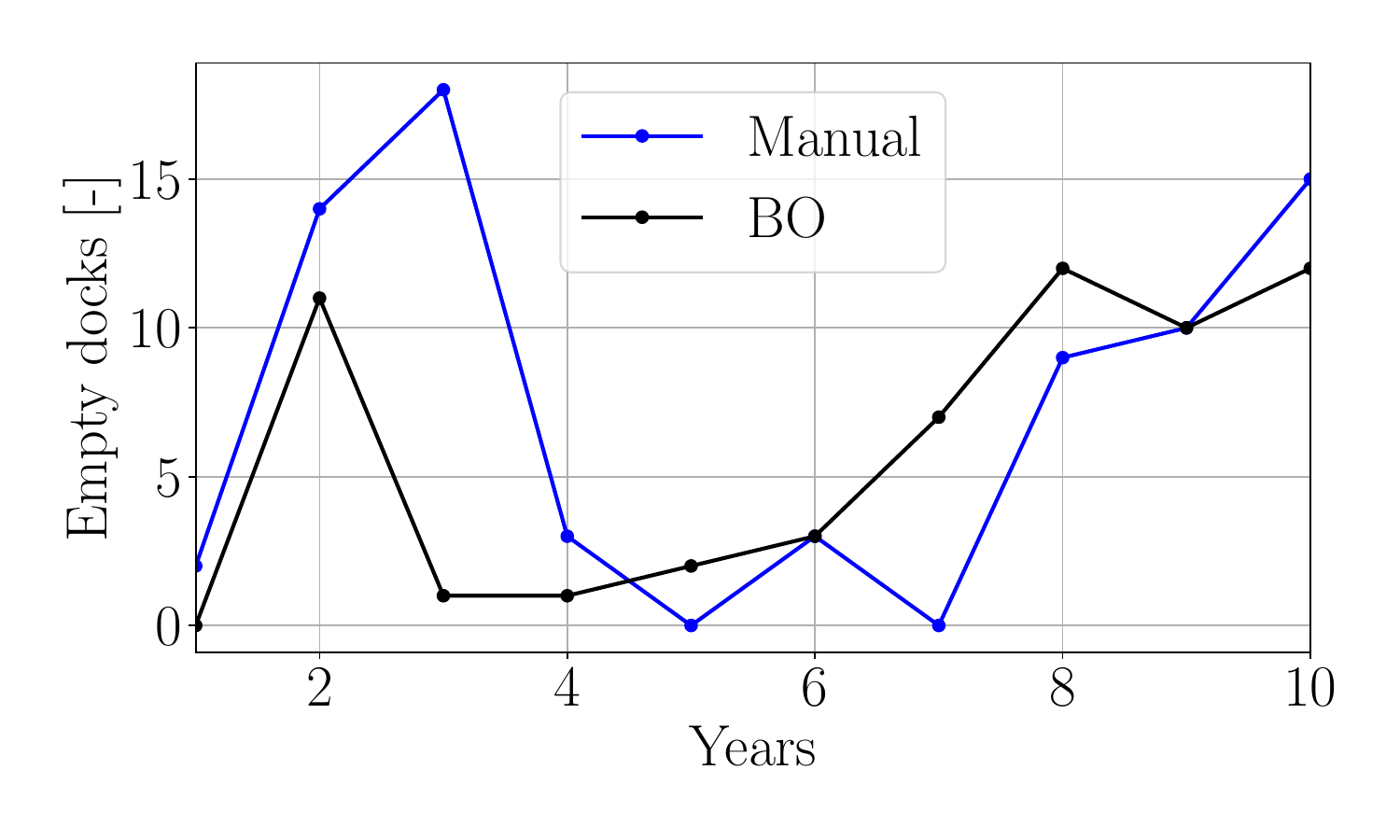}} & \subfigure[Scaling cost: manual \emph{vs} BO-tuning]{\includegraphics[width=0.4\textwidth]{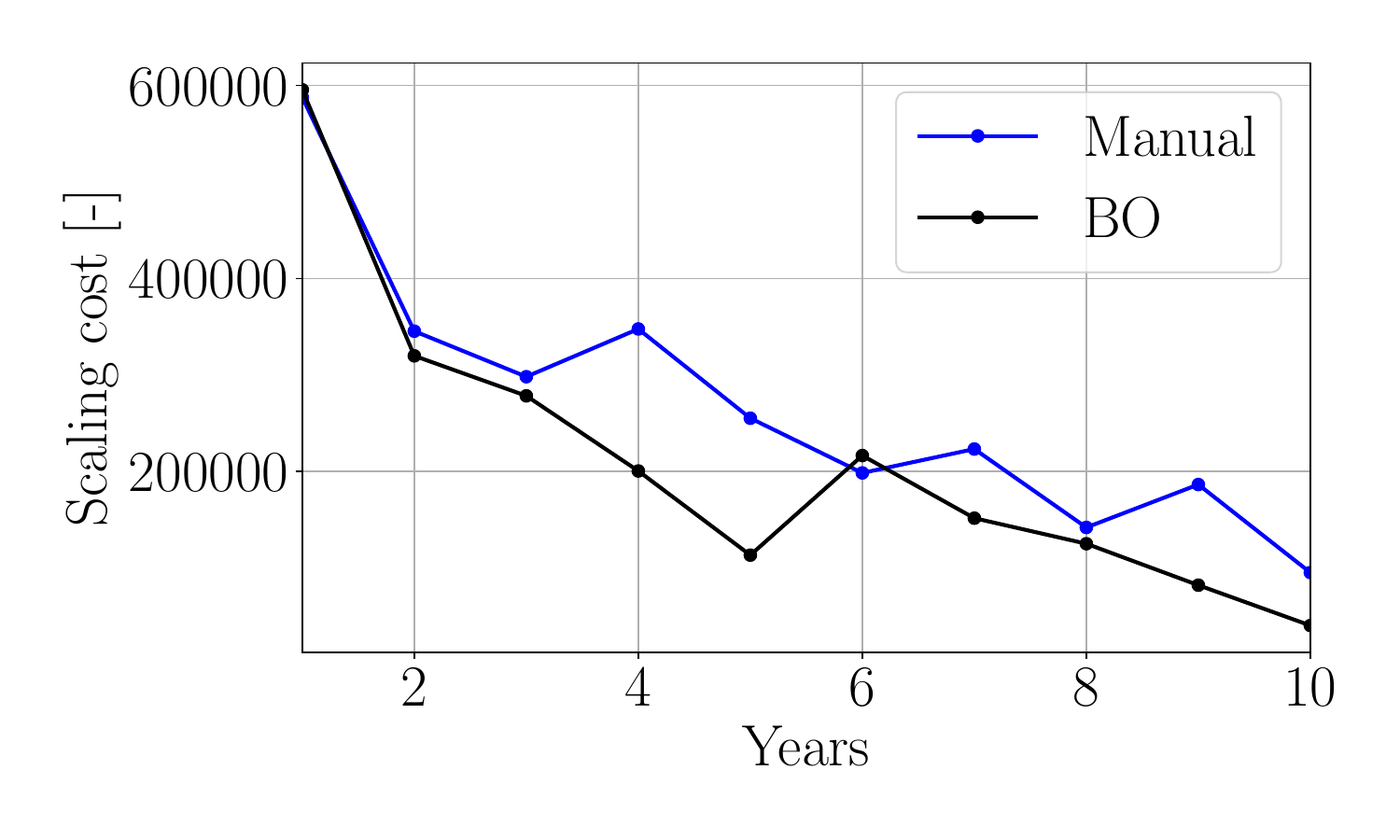}}  
    \end{tabular}
    \caption{BO effect on the critical fleet: Empty docks and scaling cost over the planning horizon.}
    \label{fig:BO_benefits1}
\end{figure}
\begin{figure}[!tb]
    \centering
    \begin{tabular}{ccc}
         \subfigure[Manual tuning: year 1]{\includegraphics[width=0.3\textwidth]{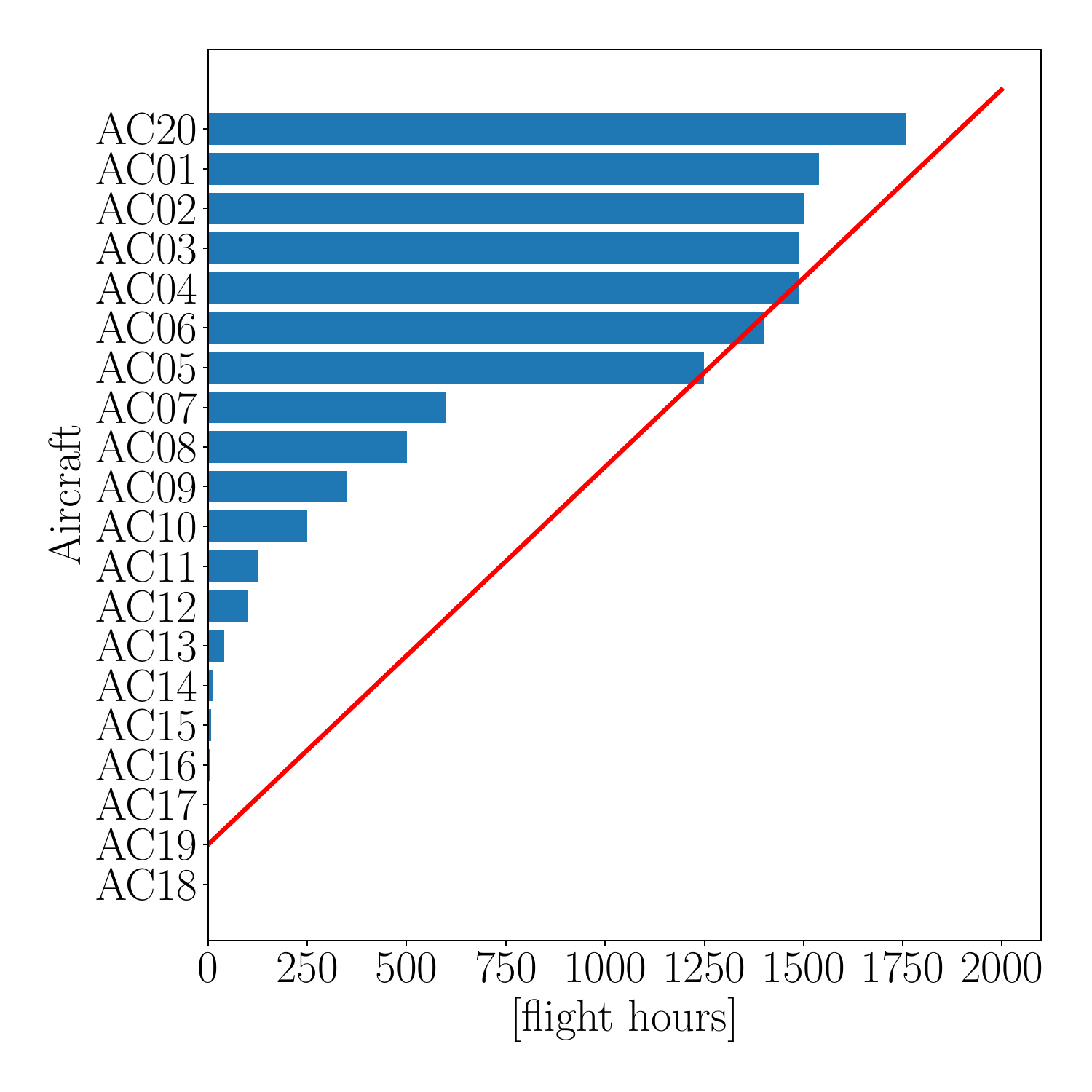}} & \subfigure[Manual tuning: year 4]{\includegraphics[width=0.3\textwidth]{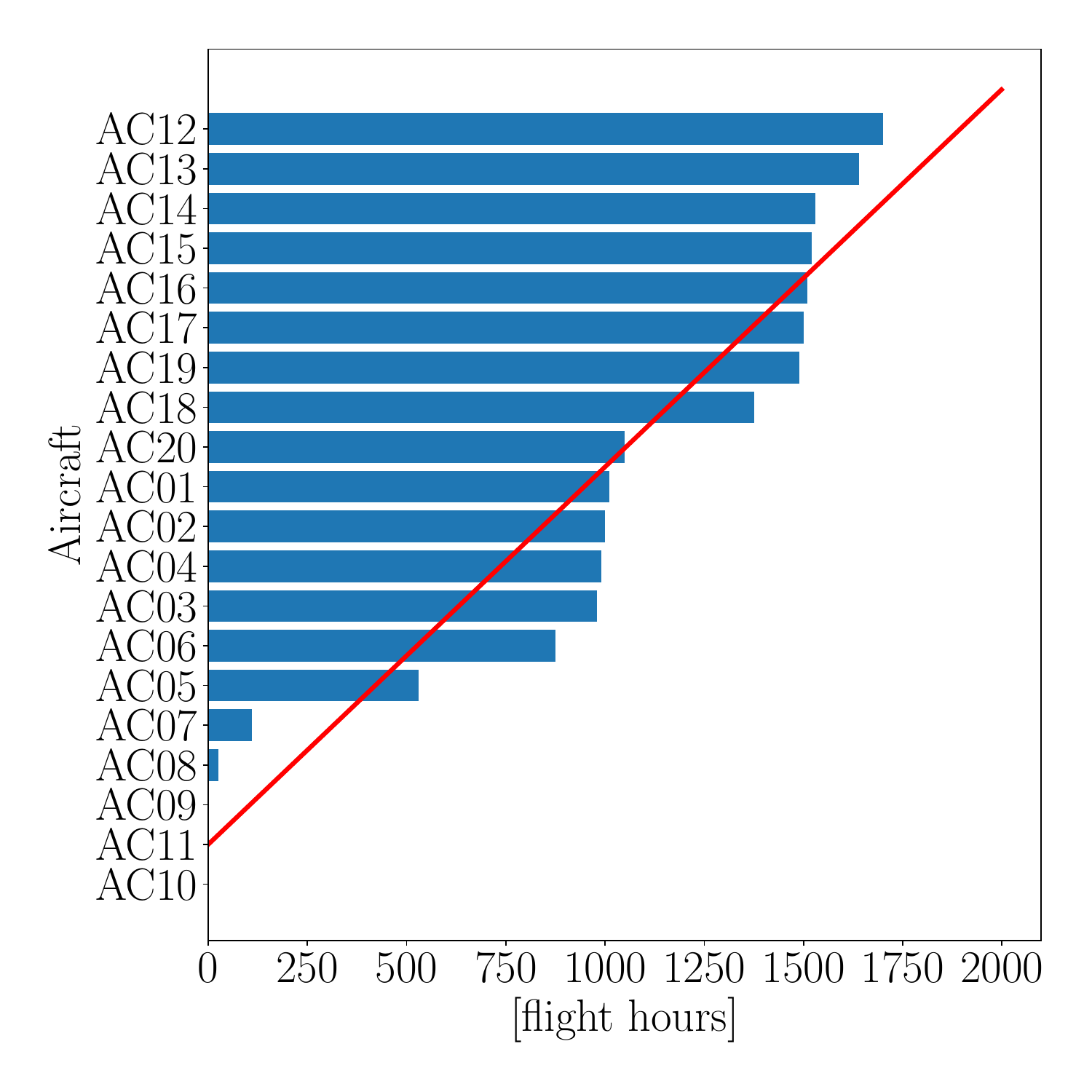}} & \subfigure[Manual tuning: year 7]{\includegraphics[width=0.3\textwidth]{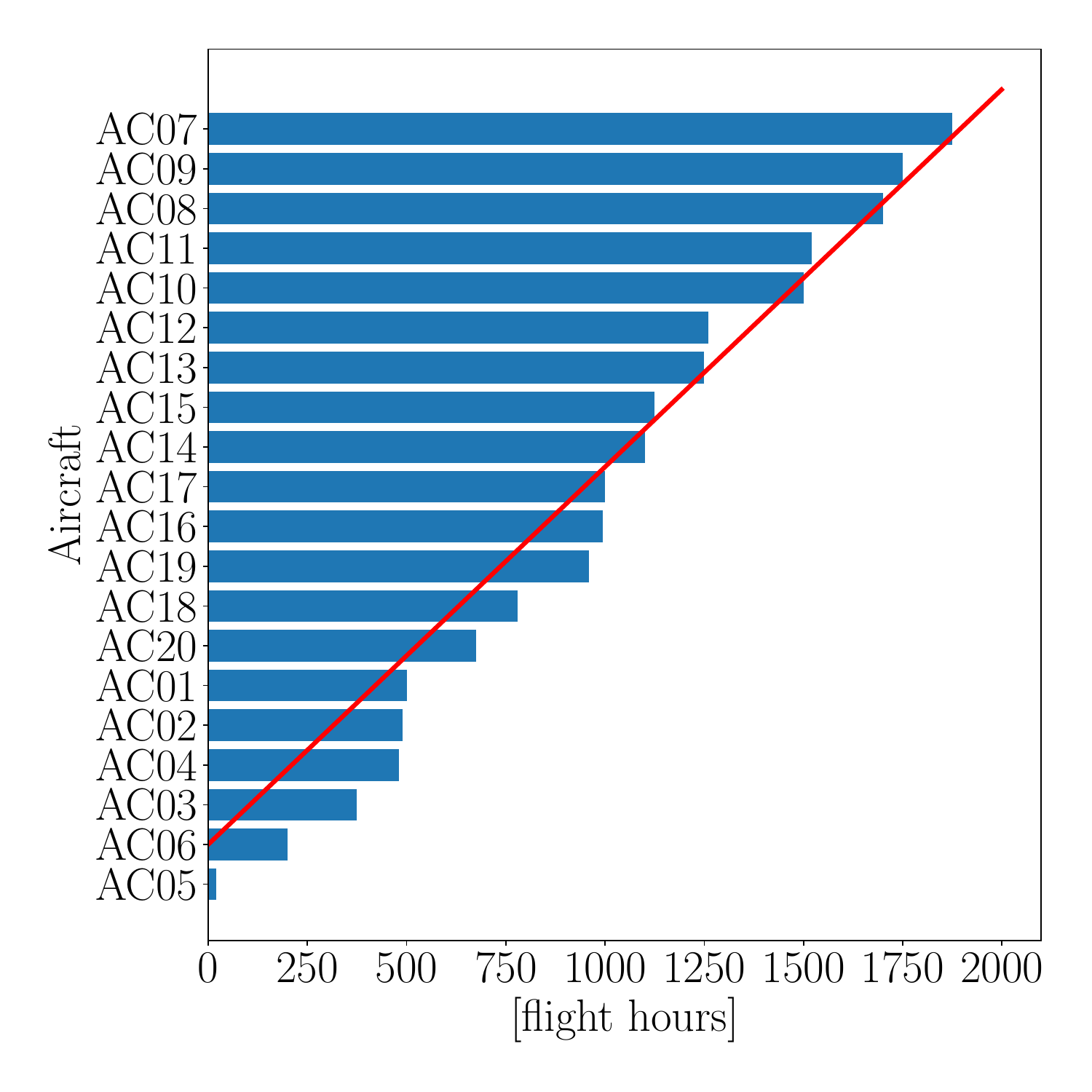}}  \\
        \subfigure[BO-based tuning: year 1]{\includegraphics[width=0.3\textwidth]{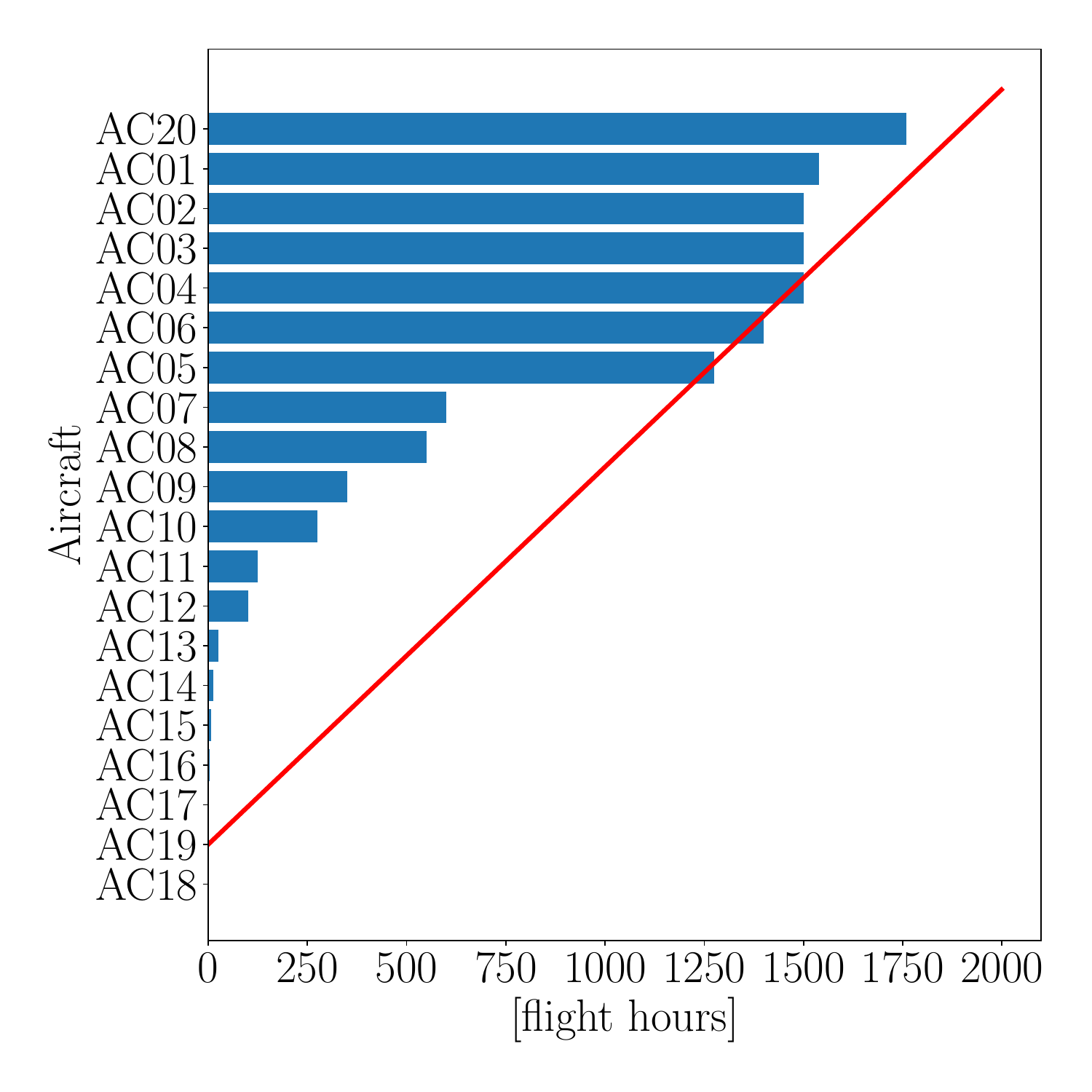}} & \subfigure[BO-based tuning: year 4]{\includegraphics[width=0.3\textwidth]{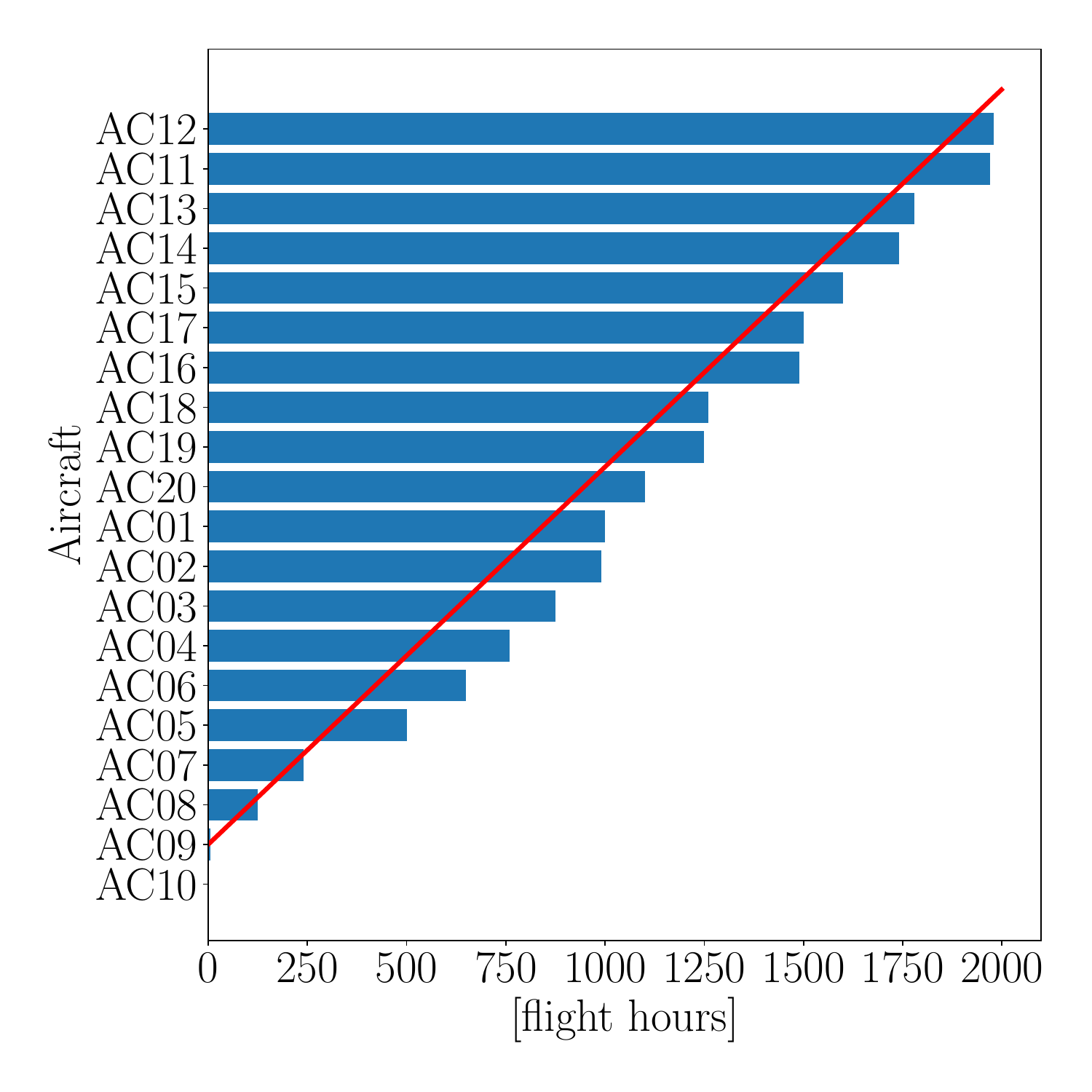}} & \subfigure[BO-based tuning: year 7]{\includegraphics[width=0.3\textwidth]{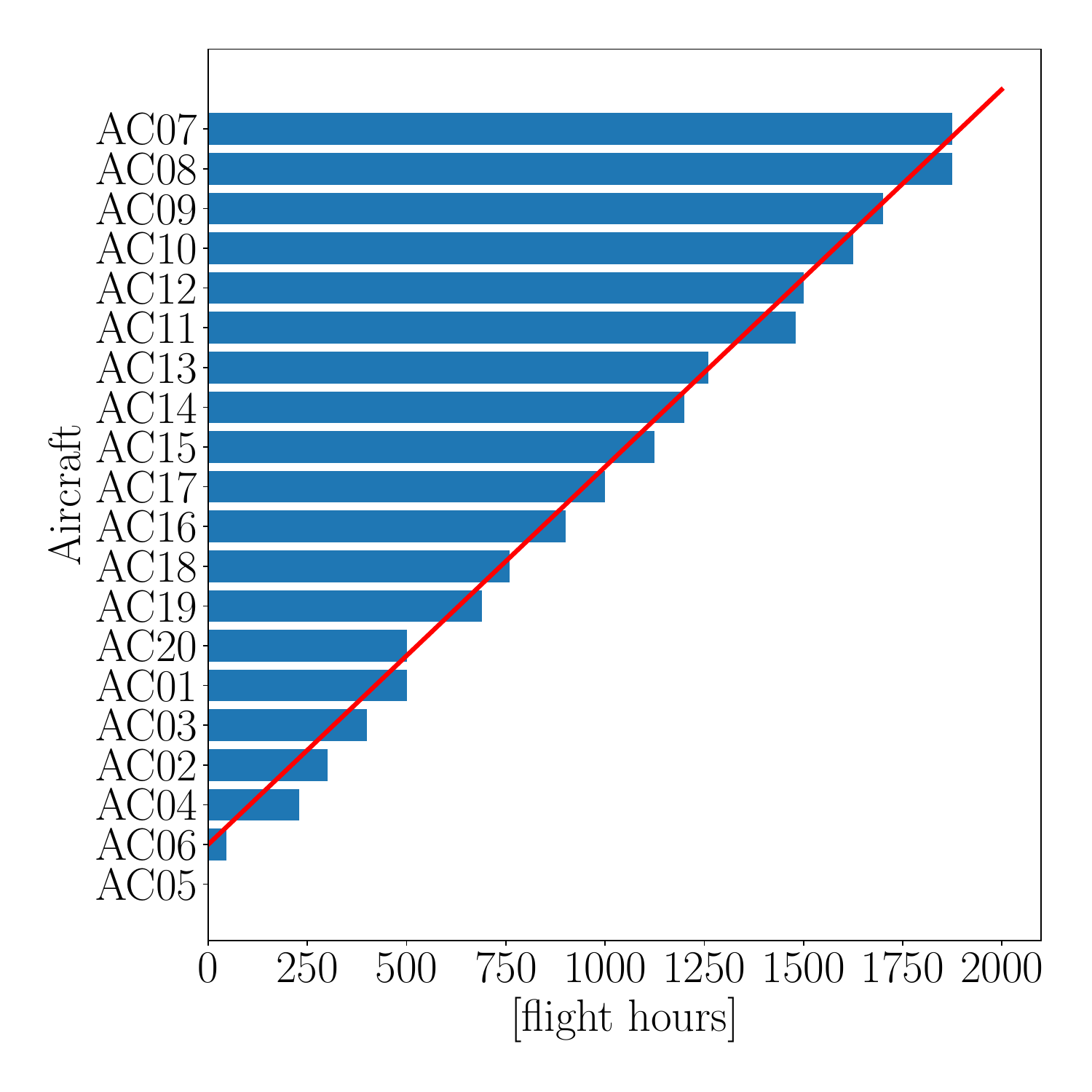}} 
    \end{tabular}
    \caption{BO effect on the critical fleet: scaling over the years with manual and BO-based tuning.}
    \label{fig:BO_benefits2}
\end{figure}
We now analyse the impact of the BO-based penalty tuning strategy presented in Section~\ref{sec:BO_hyperparameter}, focusing on the critical fleet a year-by-year planning. For the sake of visualizing the long-term benefits of adopting the automatic tuning approach, we make the planning horizon equal to $10$ years. In table \eqref{tab:weights_critical_bo} the two employed sets of weights are reported: the BO outcome vs. a trial-and-error manual tuning attempt. As shown in~\figurename{~\ref{fig:BO_benefits1}}, automatic tuning globally reduces the number of empty docks throughout the planning horizon. At the same time we observe a consistent improvements in scaling cost, as confirmed by the closer alignment between achieved and ideal scaling in~\figurename{~\ref{fig:BO_benefits2}}.

\subsection{Enhancing new aircraft handling with the hierarchical approach}\label{sec:results_hierarchy}
\begin{figure}[!tb]
    \centering
    \begin{tabular}{ccc}
         \subfigure[No-hierarchy: year 1]{\includegraphics[width=0.3\textwidth]{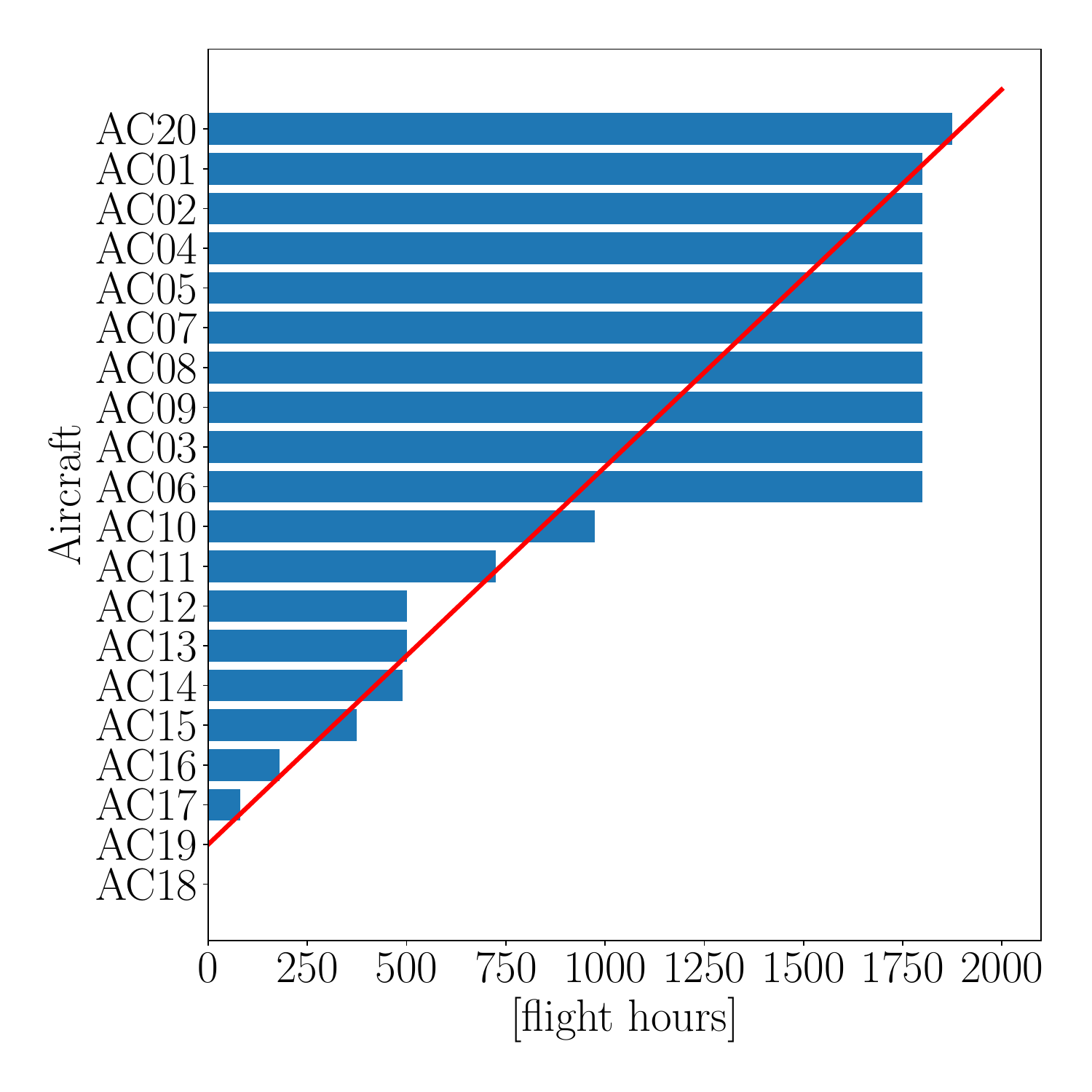}} & \subfigure[No-hierarchy: year 6]{\includegraphics[width=0.3\textwidth]{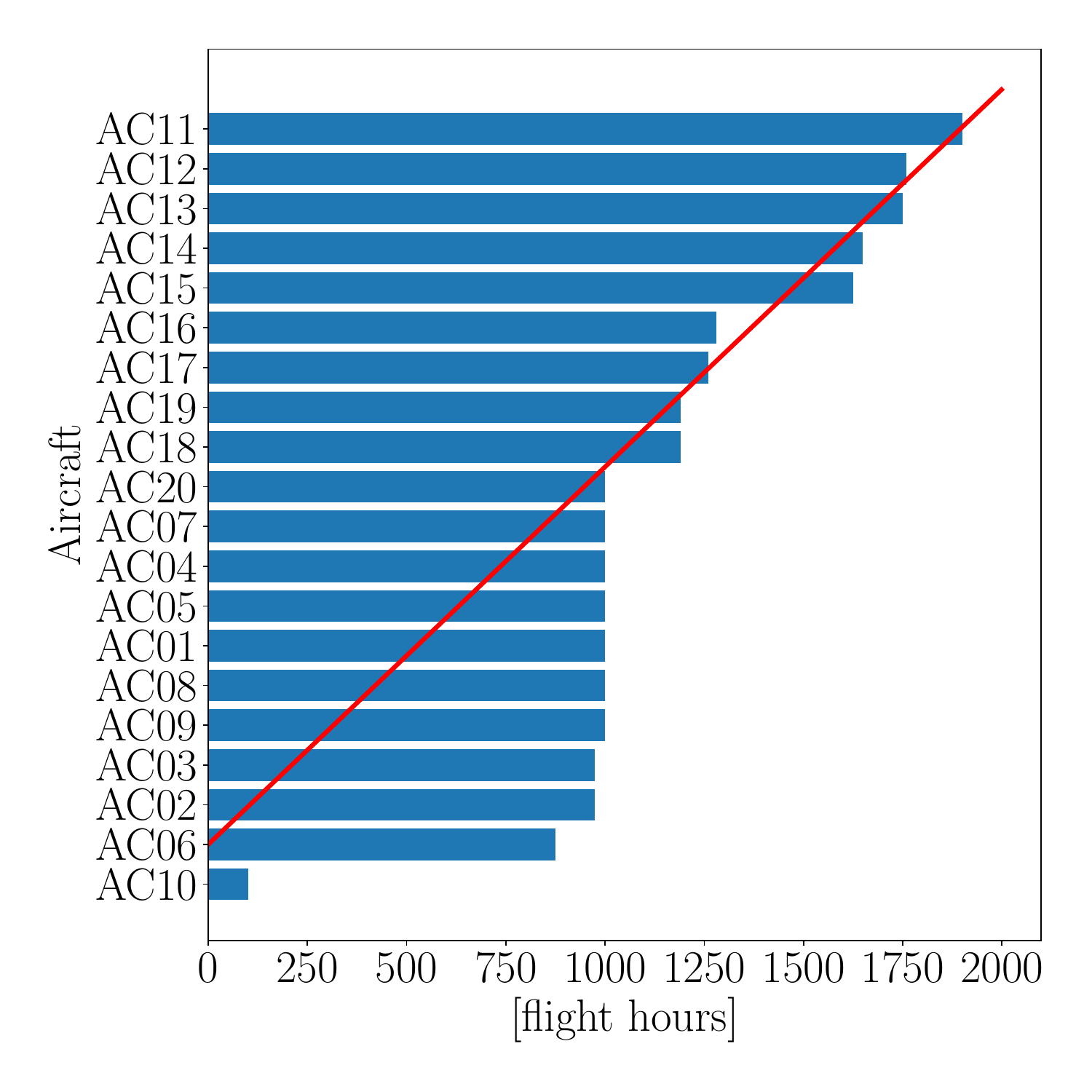}} & \subfigure[No-hierarchy: year 10]{\includegraphics[width=0.3\textwidth]{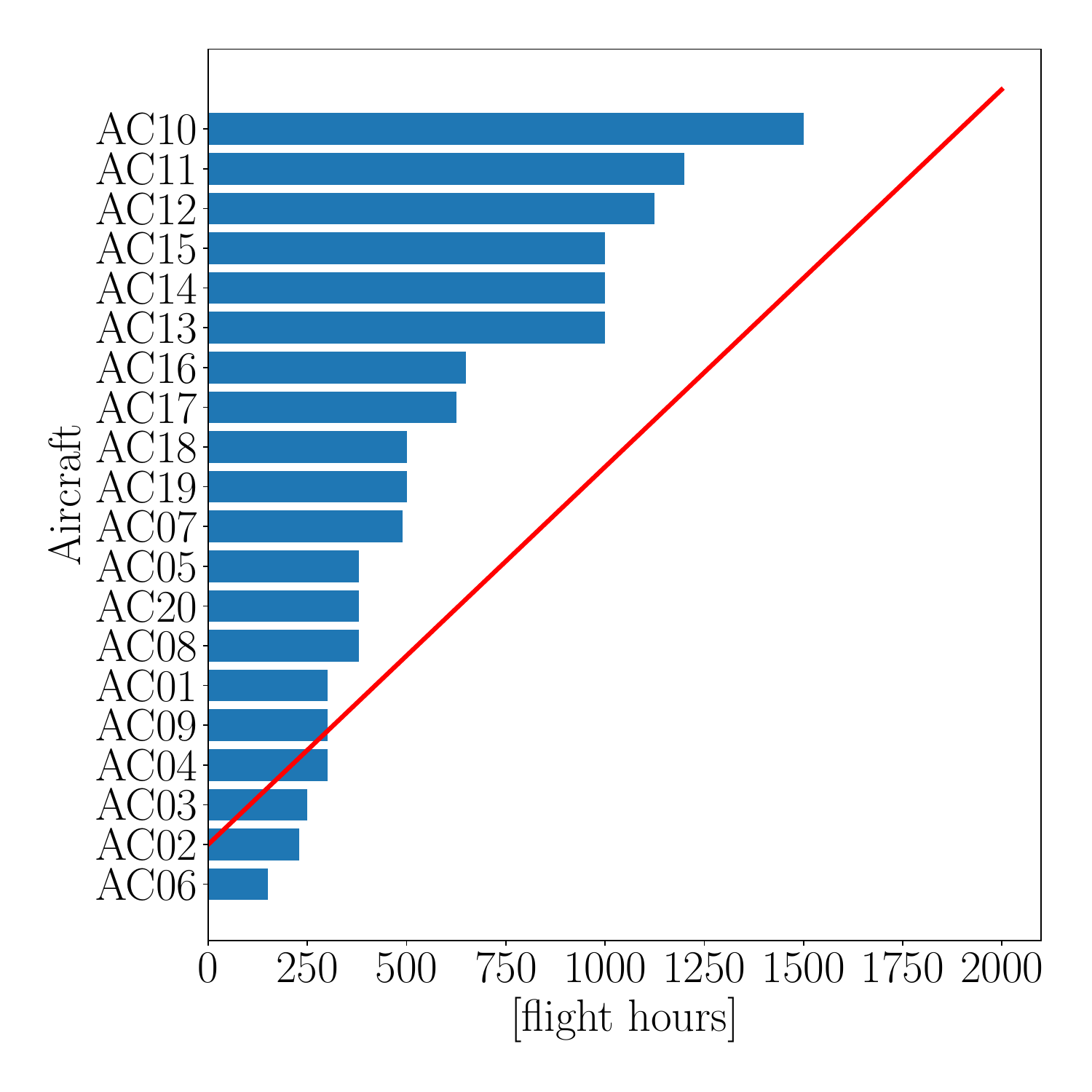}}  \\
        \subfigure[Hierarchy: year 1]{\includegraphics[width=0.3\textwidth]{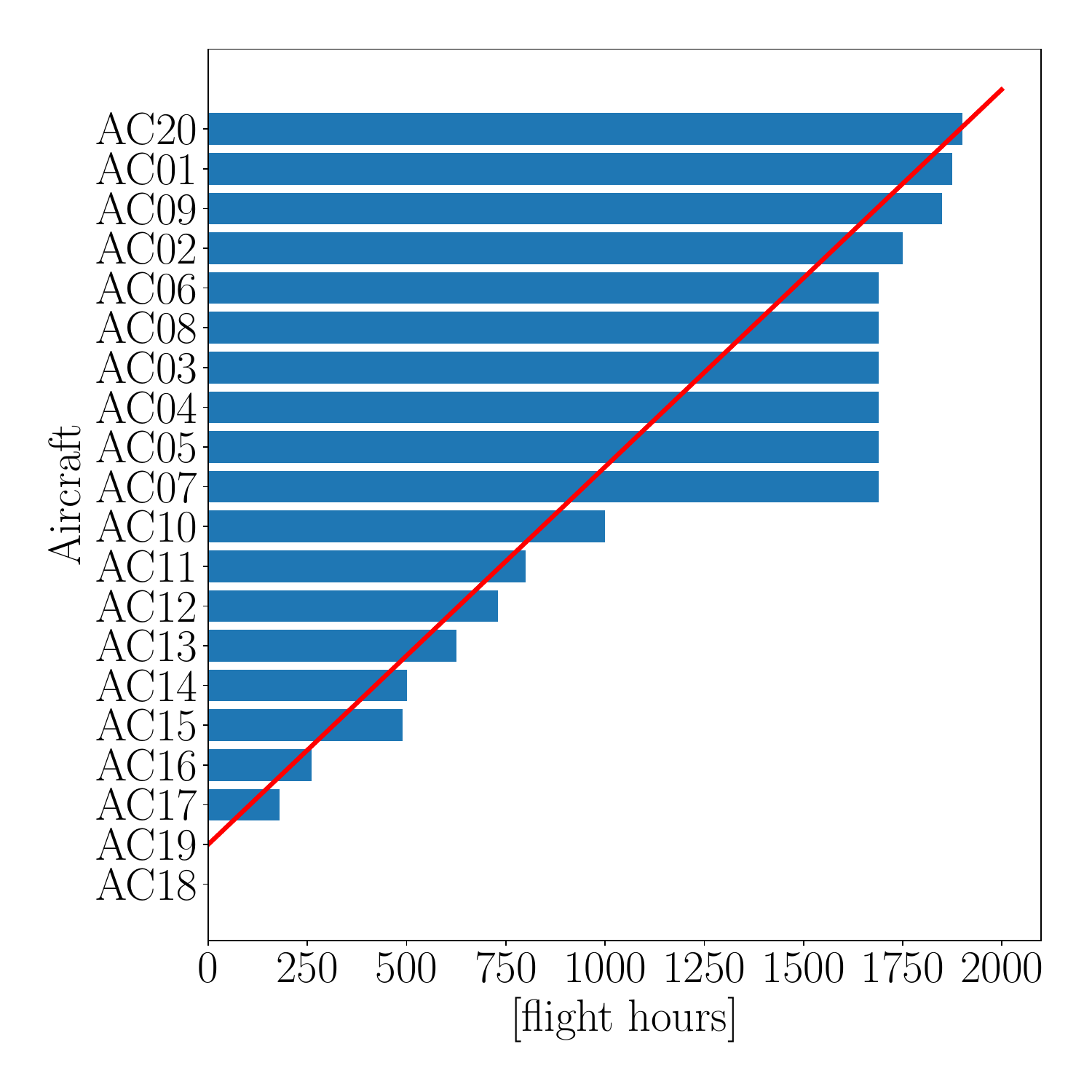}} & \subfigure[Hierarchy: year 6]{\includegraphics[width=0.3\textwidth]{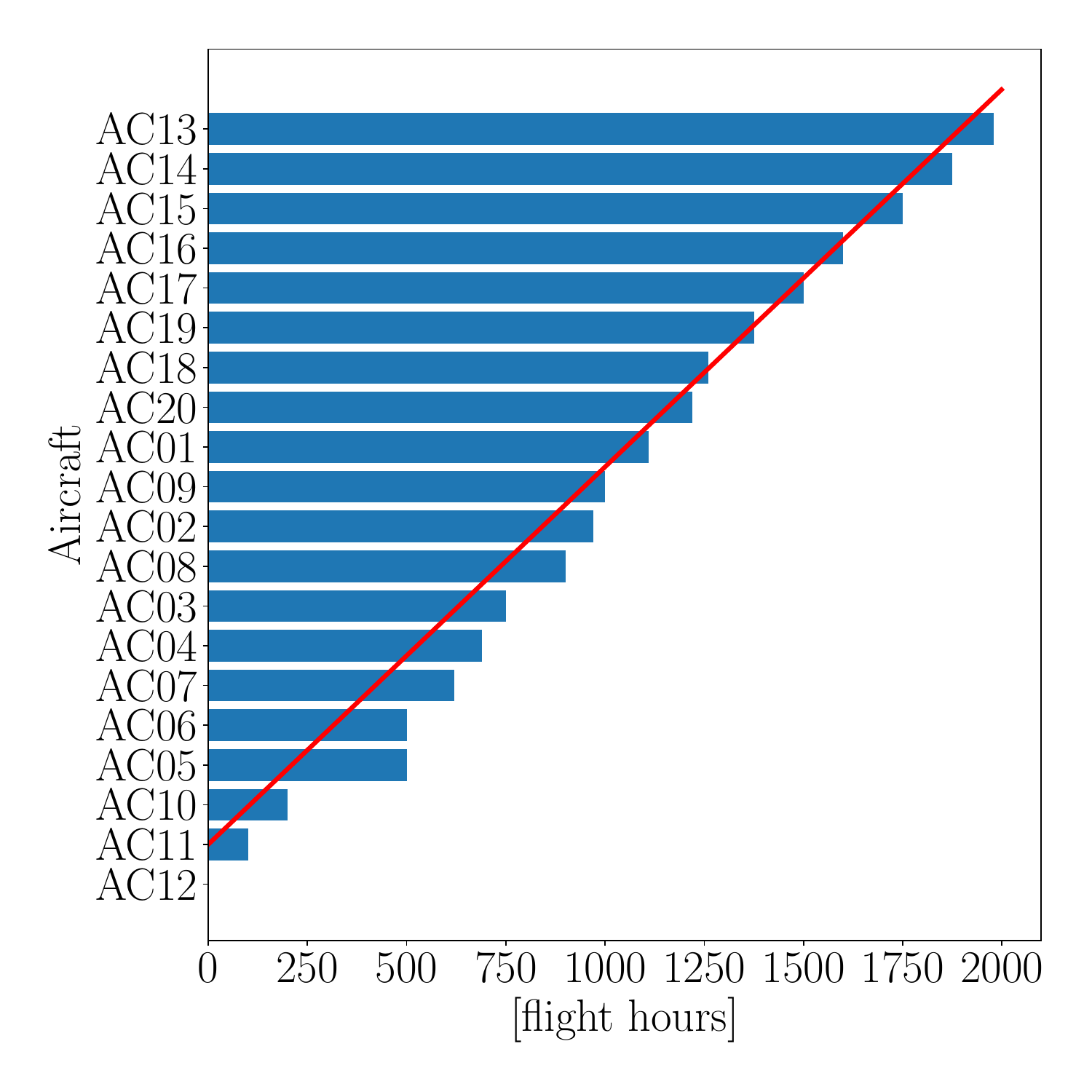}} & \subfigure[Hierarchy: year 10]{\includegraphics[width=0.3\textwidth]{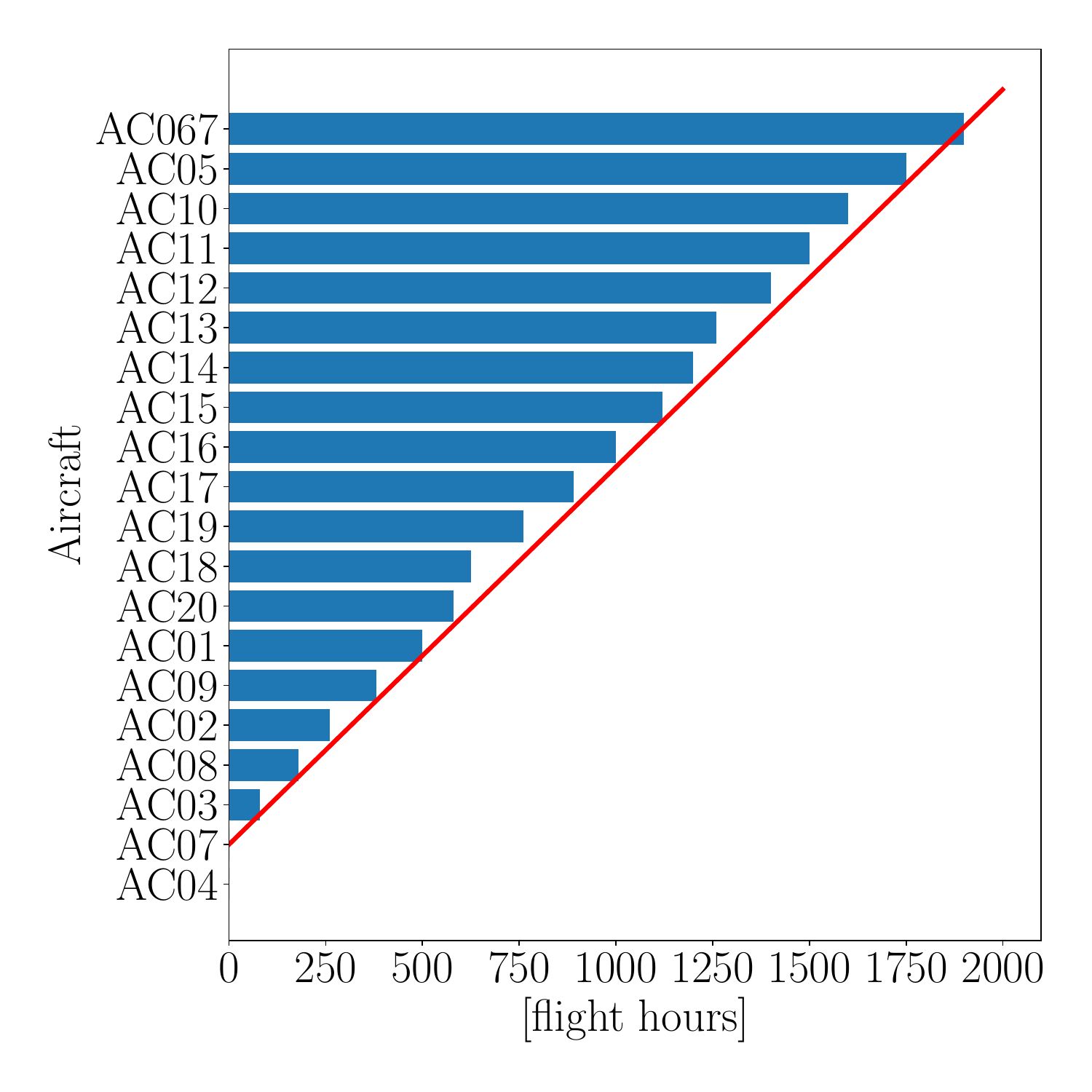}} 
    \end{tabular}
    \caption{Hierarchy effect on the new fleet: scaling obtained with and without hierarchical weights over the years.}
    \label{fig:hierarc_benefits2}
\end{figure}
Lastly, we assess the benefits of adopting the hierarchical strategy proposed in Section~\ref{sec:hierarchy} in the setting when it is likely to be more needed, i.e., when considering the new fleet. Indeed, in this case, the initial scaling does not introduce a hierarchy among aircraft on its own, which might make it more difficult for the FMP scheme to prioritize which aircraft should enter maintenance and consume FH and when. Once again, to evaluate the long-term benefits of using the hierarchical approach, we extend the planning horizon to $10$ years. 
Clearly, by \textquotedblleft artificially\textquotedblright \ enforcing a hierarchy among the aircraft we drastically reduce the surplus of FH in the fleet, while overall not decreasing its deficit. This result aligns with the scaling reported in \figurename{~\ref{fig:hierarc_benefits2}}. 
Without the hierarchy, new aircraft are treated uniformly and converge toward major inspections simultaneously, creating long-term queues. Enforcing hierarchical penalties instead spreads RFH more evenly, aligning new aircraft with ideal inspection timing and reducing queues and operational inefficiencies.

\section{Conclusions and future work}\label{sec:conclusions}
This work formulates a multi-objective, multi-year flight and maintenance planning problem for defense fleets, focusing on maximizing availability while satisfying operational constraints rather than economic objectives. Given the computational complexity of the resulting mixed-integer problem, we adopt a receding-horizon strategy to ease computation and ensure smooth transitions between yearly plans. Results show clear advantages over myopic year-by-year planning while confirming the approach’s computational feasibility.
To address the problem’s multi-objective nature, we introduce a Bayesian Optimization strategy to auto-tune cost penalties. Results on a fleet with a \emph{critical} initial scaling highlight the advantages of automatic tuning over manual selection and its potential to reduce user burden. Finally, we propose custom penalty coefficients' definitions to handle cases where multiple aircraft share similar residual flight hours, mitigating inspection queues and improving overall fleet efficiency. Results on a benchmark fleet that has just received a new batch of aircraft demonstrate the benefits of aircraft-specific weighting, highlighting its potential to tailor the FMP model to fleet-specific needs.
Despite these benefits, the proposed scheme can be further improved. Currently, it assumes a cyclic inspection framework, which may not reflect all military fleet operations. Future works will thus be devoted to extend the approach to acyclic and calendar-dependent inspections, as well as to maintenance anticipations and delays commonly encountered in practice. Another key direction is to enhance robustness against unexpected and stochastic events that may disrupt effective flight and maintenance planning if not explicitly modelled.

\section*{Acknowledgements}

This work was supported by Aeronautica Militare Italiana under the P.A.F.A.M. project.

\section*{Declaration of Generative AI and AI-assisted technologies in the writing process}
During the preparation of this work the authors used ChatGPT in order to improve readability. After using this tool, the authors reviewed and edited the content as needed and take full responsibility for the content of the publication.

\bibliographystyle{elsarticle-harv}
\bibliography{references}

@book{bernardo2009bayesian,
  title={Bayesian theory},
  author={Bernardo, Jos{\'e} M and Smith, Adrian FM},
  volume={405},
  year={2009},
  publisher={John Wiley \& Sons}
}

@article{victoria2021automatic,
  title={Automatic tuning of hyperparameters using Bayesian optimization},
  author={Victoria, A Helen and Maragatham, Ganesh},
  journal={Evolving Systems},
  volume={12},
  number={1},
  pages={217--223},
  year={2021},
  publisher={Springer}
}

@article{pelikan2005bayesian,
  title={Bayesian optimization algorithm},
  author={Pelikan, Martin and Pelikan, Martin},
  journal={Hierarchical Bayesian optimization algorithm: toward a new generation of evolutionary algorithms},
  pages={31--48},
  year={2005},
  publisher={Springer}
}

@article{FlottaGreca2,
author={Kozanidis, George and George, Liberopoulos and Christos, Pitsilkas},
title = {Flight and maintenance planning of military aircraft for maximum fleet availability.},
journal = {Military Operations Research},
year = {2010},
pages = {53-73}
}

@article{FlottaGreca3,
author={Gavranis, Andreas and George, Kozanidis},
title = {Mixed integer biobjective quadratic programming for maximum-value minimum-variability fleet availability of a unit of mission aircraft.},
journal = {Computers and Industrial Engineering},
year = {2017},
pages = {13-29}
}

@article{FlottaGreca4,
author={Gavranis, Andreas and George, Kozanidis},
title = {An exact solution algorithm for maximizing the fleet availability of a unit of aircraft subject to flight and maintenance requirements.},
journal = {European Journal of Operational Research},
volume={242},
number={2},
year = {2015},
pages = {631-643}
}

@article{Peschiera2,
author={Peschiera, Franco and Battaïa, Olga and Haït1, Alain and Dupin, Nicolas},
title = {Long term planning of military aircraft flight and maintenance operations.},
journal = {arXiv preprint arXiv:2001.09856},
year = {2020},
}

@article{grecoMultiObj,
author={Kozanidis, George},
title = {A multiobjective model for maximizing fleet availability under the presence of flight and maintenance requirements.},
journal = {Journal of Advanced Transportation},
year = {2006},
pages = {155-182},
volume={434},
number={2},
}

@article{Mattila,
author={Mattila, Ville and Kai, Virtanen},
title = {Maintenance scheduling of a fleet of fighter aircraft through multi-objective simulation-optimization.},
journal = {Simulation},
year = {2014},
pages = {1023-1040},
volume={90},
number={9},
}

@article{NavalResearchGreco,
author={Kozanidis, George and Andreas, Gavranis and Eftychia, Kostarelou},
title = {Mixed integer least squares optimization for flight and maintenance planning of mission aircraft.},
journal = {Naval Research Logistics (NRL) },
year = {2012},
pages = {212-229},
volume={59},
number={3‐4},
}

@article{Peschiera,
  title={A novel solution approach with ML-based pseudo-cuts for the Flight and Maintenance Planning problem.},
  author={Peschiera, Franco and Dell, Robert and Royset, Johannes and Haït1, Alain and Dupin, Nicolas and Battaïa, Olga},
  journal={OR Spectrum},
  volume={43},
  number={3},
  pages={635-664},
  year={2021},
}

@article{Verhoeff,
  title={Maximizing operational readiness in military aviation by optimizing flight and maintenance planning.},
  author={Verhoeff, M. and W. J. C., Verhagen and Richard, Curran},
  journal={Transportation Research Procedia},
  volume={10},
  pages={941-950},
  year={2015},
}

@article{Safaei,
  title={Workforce-constrained maintenance scheduling for military aircraft fleet: a case study.},
  author={Safaei, Nima and Dragan, Banjevic and Andrew KS, Jardine},
  journal={Annals of Operations Research},
  volume={186},
  number={1},
  pages={295-316},
  year={2011},
}

@article{biro_civil,
  title={Aircraft and maintenance scheduling support, mathematical insights and a proposed interactive system},
  author={Biró, M. and Simon, I. and Tánczos, C.},
  journal={Journal of Advanced Transportation},
  volume={26},
  pages={121-130},
  year={1992},
}

@article{clarke_civil,
  title={The aircraft rotation problem},
  author={Clarke, L. W. and Johnson, E. L. and Nemhauser, G. L. and Zhu, Z.},
  journal={Annals of Operations Research},
  volume={69},
  pages={33-46},
  year={1997},
}

@article{gopalan_civil,
  title={Mathematical models in airline schedule planning: a survey},
  author={Gopalan, R. and Talluri, K. L.},
  journal={Annals of Operational Research},
  volume={76},
  pages={155-185},
  year={1998},
}

@article{moudani_civil,
  title={A dynamic approach for aircraft assignment and maintenance scheduling by airlines.},
  author={El Moudani, W. and Mora-Camino, F.},
  journal={Journal of Air Transport Management},
  volume={6},
  number={},
  pages={233-237},
  year={2000},
}

@article{siriam_civil,
  title={An optimization model for aircraft maintenance scheduling and reassignment.},
  author={Sriram, C. and Haghani, A.},
  journal={Transportation Research Part A: Policy and Practice},
  volume={37},
  number={},
  pages={29-48},
  year={2003},
}

@article{cohn_civil,
  title={Improving crew scheduling by incorporating key maintenance routing decisions.},
  author={Cohn, A. M. and Barnhart, C.},
  journal={Operation Research},
  volume={51},
  number={},
  pages={387-396},
  year={2003},
}

@article{yan_civil1,
  title={Air cargo fleet routing and timetable setting with multiple on-time demands.},
  author={Yan, S. and Chen, S.-C. and Chen, C.-H.},
  journal={Transportation Research Part E: Logistics and Transportation Review},
  volume={42},
  pages={409–430},
  year={2006},
}

@article{yan_civil2,
  title={A flight scheduling model for Taiwan airlines under market competitions},
  author={Yan, S. and Tanga, C.-H. and Leea, M.-C.},
  journal={Omega},
  volume={35},
  number={},
  pages={61–74},
  year={2007},
}

@article{ahire_civil,
  title={Workforce-constrained preventive maintenance scheduling using evolution strategies},
  author={Ahire, S. and Greenwood, G. and Gupta, A. and Terwilliger, M.},
  journal={Decision Sciences Journal},
  volume={31},
  pages={833–859},
  year={2000},
}

@article{dijkstra_civil,
  title={Planning the size and organisation of KLM’s aircraft maintenance personnel},
  author={Dijkstra, M. C. and Kroon, L. G. and Salomon, M. and Van Nunen, J. and Van Wassenhove, L. N.},
  journal={Interfaces},
  volume={24},
  number={},
  pages={47–58},
  year={1994},
}

@article{radosavljevic_militar,
  title={Assigning Fighter Plane Formations to Enemy Aircraft using Fuzzy Logic},
  author={Radosavljevic, Z. and Babic, O.},
  journal={Transportation Planning and Technology},
  volume={23},
  pages={353-368},
  year={2000},
}

@article{kurokawa_militar,
  title={Air Transportation Planning using Neural Networks as an Example of the Transportation Squadron in the Japan Air Self-Defense Force},
  author={Kurokawa, T. and Takeshita K.},
  journal={Systems and Computers in Japan},
  volume={35-12},
  pages={1223-1232},
  year={2004},
}

@article{yeung_militar,
  title={Mission Assignment and Maintenance Scheduling for Multi-State Systems},
  author={Yeung, T.G. and Cassady C.R. and Pohl, E.A.},
  journal={Military Operations Research},
  volume={12-1},
  number={},
  pages={19-34},
  year={2007},
}

@article{sgaslik_militar,
  title={Planning German Army Helicopter Maintenance and Mission Assignment},
  author={Sgaslik, A.},
  journal={MSc Thesis, Naval Postgraduate School, Monterey, USA},
  year={1994}
}

@article{pippin_militar,
  title={Allocating Flight Hours to Army Helicopters},
  author={Pippin, B.W.},
  journal={MSc Thesis, Naval Postgraduate School, Monterey, CA, USA,},
  volume={},
  number={},
  pages={},
  year={1998},
}

@article{usa_manual,
  title={Field Manual No. 304.500: Army Aviation Maintenance (Appendix D: Maintenance Management Tools)},
  author={U.S. DoA, United States Department of the Army},
  journal={},
  year={2000},
}

@article{BO_ref,
title = {Hyperparameter Optimization for Machine Learning Models Based on Bayesian Optimization},
journal = {Journal of Electronic Science and Technology},
volume = {17},
number = {1},
pages = {26-40},
year = {2019},
author = {Jia Wu and Xiu-Yun Chen and Hao Zhang and Li-Dong Xiong and Hang Lei and Si-Hao Deng},
}

@article{rolling_horiz_ref1,
  title={A theory of rolling horizon decision making},
  author={Sethi, Suresh and Sorger, Gerhard},
  journal={Annals of Operations Research},
  volume={29},
  year={1991},
}

@article{rolling_horiz_ref2,
author = {Funda Sahin and Arunachalam Narayanan and E. Powell Robinson and},
title = {Rolling horizon planning in supply chains: review, implications and directions for future research},
journal = {International Journal of Production Research},
volume = {51},
number = {18},
pages = {5413--5436},
year = {2013}
}

@article{bemporad_bo_ref,
  title={Performance-Oriented Model Learning for Data-Driven MPC Design},
  author={Piga, Dario and Forgione, Marco and Formentin, Simone and Bemporad, Alberto},
  journal={IEEE Control Systems Letters},
  year={2019}
}

\end{document}